\documentclass[preprint,12pt]{elsarticle}

\usepackage{amssymb}
\usepackage{longtable}
\usepackage{pdflscape}
\usepackage[colorlinks=true,allcolors=blue]{hyperref}
\usepackage{smartdiagram}
\usepackage{threeparttable}
\usepackage{ragged2e}
\usepackage{scalerel,stackengine}
\usepackage{subcaption}
\usepackage[labelformat=parens,labelsep=quad,skip=3pt]{caption}
\usepackage{booktabs, caption, makecell}
\usepackage{xcolor}

\usepackage{listings}
\usepackage{colortbl}
\usepackage{array}
\newcolumntype{L}[1]{>{\raggedright\arraybackslash}p{#1}}
\newcolumntype{C}[1]{>{\centering\arraybackslash}p{#1}}
\newcolumntype{R}[1]{>{\raggedleft\arraybackslash}p{#1}}
\usepackage{multirow}
\usepackage{dcolumn,longtable,booktabs}
\usepackage[most]{tcolorbox}
\usepackage{xr}
\definecolor{Gray}{gray}{0.9}
\usepackage{tcolorbox}

\journal{Information and Software Technology}

\begin{document}

\begin{frontmatter}

\title{Community Smells - The Sources of Social Debt: \\ A Systematic Literature Review}

\author[1,2]{Eduardo Caballero-Espinosa}
\ead{eduardo.caballero@utp.ac.pa}
\author[3]{Jeffrey C. Carver}
\ead{carver@cs.ua.edu}
\author[4]{Kimberly Stowers}
\ead{kim.stwrs@gmail.com}

\address[1]{Center for Research, Development, and Innovation in Information and Communication Technology (CIDITIC) - Technological University of Panama}
\address[2]{Computing Systems Engineering Department - Technological University of Panama}
\address[3]{Department of Computer Science, The University of Alabama, Tuscaloosa, AL, USA}
\address[4]{Tim Fletcher, Co., San Jose, CA, USA}

\begin{abstract}
   \textit{Context}:
Social debt describes the accumulation of unforeseen project costs (or potential costs) from sub-optimal software development processes.
Community smells are sociotechnical anti-patterns and one source of social debt.
Because community smells impact software teams, development processes, outcomes, and organizations, we to understand their impact on software engineering.

\textit{Objective}:
To provide an overview of community smells in social debt, based on published literature, and describe future research.

\textit{Method}:
We conducted a systematic literature review (SLR) to identify properties, understand origins and evolution, and describe the emergence of community smells.
This SLR explains the impact of community smells on teamwork and team performance.

\textit{Results}:
We include 25 studies.
Social debt describes the impacts of poor socio-technical decisions on work environments, people, software products, and society.
For each of the 30 community smells identified as sources of social debt, we provide a detailed description, management approaches, organizational strategies, and mitigation effectiveness. 
We identify five groups of management approaches: organizational strategies, frameworks, models, tools, and guidelines.
We describe 11 common properties of community smells.
We develop the \textit{Community Smell Stages Framework} to concisely describe the origin and evolution of community smells.
We then describe the causes and effects for each community smell.
We identify and describe 8 types of causes and 11 types of effects related to the community smells.
Finally, we provide 8 comprehensive Sankey diagrams that offer insights into threats the community smells pose to teamwork factors and team performance.

\textit{Conclusion}:
Community smells explain the influence work conditions have on software developers.
The literature is scarce and focuses on a small number of community smells.
Thus, the community smells still need more research.
This review helps by organizing the state of the art about community smells.
Our contributions provide motivations for future research and provide educational material for software engineering professionals.
\end{abstract}


\end{frontmatter}

\section{Introduction}
    Historically, software developers, customers, end-users, and other stakeholders have viewed the software development process as an \textit{entirely technical activity}.
To produce high-quality software, organizations commonly hire qualified professionals, provide adequate infrastructure, and use efficient project management approaches.
A number of methods, tools, standards, and other technical approaches support these activities~\cite{Garousi_intro_100HighlycitedpapersSE:2016}.
However, human involvement in the software development process has led stakeholders to recognize that it is also \textit{social activity}~\cite{Dittrich_socialSoftware:2002} in which the interactions among people are central.
Therefore, software development is a socio-technical activity where both social factors and technical factors are essential for success.
This link between social and technical factors also helps explain Conway's law~\cite{Conway_law:1968}, which states that design systems match the communication structures of the organizations that produce it.

Together the social and technical decisions shape the work environment, including the policies, procedures, and tools, that ultimately affect developers' psychology.
Imagine a situation in which an organization allocates developers to three teams who work together to develop a mobile application.
Then, in the middle of the process, the project leader reorganizes the teams (a social change) and introduces a new development methodology (a technical change).
The developers' reactions to these changes can be mistrust, irritation, and miscommunication.
When these reactions persist, they increase cost and introduce \textit{social debt}.

\textit{Social debt} describes unforeseen project costs (or potential project costs) related to sub-optimal software development teams~\cite{Tamburri-etal_whatSDinSE:2013}.
To explain social debt, researchers focus on identifying poor socio-technical decisions and their impact on software development organizations.
\textit{Community smells} describe connections between the poor socio-technical decisions that shape work environments and their adverse effects on individuals and software development teams.
Tamburri et al.~\cite{Tamburri-etal_socialDebtSEinsights:2015} define community smells as ``\textit{sets of organizational and social circumstances, having implicit causal relations}.''
Community smells primarily produce powerful emotions, stress social interactions, affect team performance, and decrease software quality.
Thus, as long as they remain in place, community smells result in additional project cost.

The concept of \textit{community smells} is similar to the Technical Debt (TD) concepts of \textit{code smells} and \textit{architectural smells}.
Code smells are symptoms of poor or postponed correct implementation decisions~\cite{Fowler_refactoring:1999}.
In the end, they are evidence of structural problems in the source code, e.g., spaghetti code~\cite{Brown_etal_AntiPatternsRefactoringInCrisis:1998}.
Architectural smells are architectural decisions that negatively impact the systems' internal quality, e.g., applying design abstractions at the wrong level of granularity~\cite{Garcia_etal_IdentifyingArchitecturalSmells:2009}.
Both code smells and architectural smells represent factors that contribute to TD~\cite{SHARMA_softwareSmells:2018}.

Although poor decisions are common triggers for all three types of smells (code, architecture, and community), these smell types differ in how they impact the software development process.
Architectural smells, which focus on wrong architectural decisions that impact the software design and code smells, which result from decisions that negatively affect code quality, both focus on technical aspects of the software.
Conversely, community smells also include the social aspects of software by focusing on poor decisions that shape software development work settings and influence people's behavior.
Therefore, we excluded architectural smells and code smells from our SLR's target topics since they show clear differences from community smells in the elements they impact.

Two motivations drive the design of our systematic literature review (SLR).
The first is to provide an overall picture of community smells and social debt to researchers and professionals in software engineering.
So they can better understand the concepts and characteristics of community smells.

Also, research on community smells has basically focused on four community smells, \textit{Organizational Silo}, \textit{Black Cloud}, \textit{Lone Wolf}, and \textit{Radio Silence}/\textit{Bottleneck}, which have impact in real scenarios~\cite{Palomba-etal_beyondTechnicalAspects:2018, Tamburri-etal_exploringCSsOS:2019, Catolino-etal_genderDiversityWomen:2019, Palomba_etal_UnderstandingCSsVariability:2021, DeStefano_etal_SplicingCommunityPatternsSmells:2020, Catolino_etal_RefactoringCommunitySmells:2020, Catolino_etal_GenderDiversityCommunitySmellsInsightsTrenches:2019}.
These studies analyzed these community smells and their impact on coordination, communication, and cooperation, which are critical factors for successful teamwork in software development teams.
However, the more general literature on teamwork provides six additional factors for good teamwork like composition and cognition~\cite{Salas-etal_teamConsiderations:2015, Bell-etal_teamCompositionABCsteamwork:2018}.
Therefore, we are extending the previous group of community smells and teamwork factors and conducting a fine-grained analysis to relate them.

Based on our motivations, the main goal of this SLR is to \textbf{perform a comprehensive review of the community smells as sources of social debt, define the connections between community smells and factors for effective teamwork, and identify opportunities for additional research on community smells}.
The key contributions of this study are:
\begin{itemize}
    \item The first secondary study on community smells as the sources of social debt in the context of software engineering;
    \item A description of the properties of community smells;
    \item A framework for the origin and evolution of community smells in software development teams and organizations; and
    \item A mapping of the theoretical connection between the community smells and critical factors for effective teamwork.
    \item Educational material (e.g., definitions, management approaches, and properties) for software engineering professionals from different settings, including academia, open-source projects, and closed-source software projects.
\end{itemize}

\section{Background}
    This section provides background information on social debt and community smells.
Then it describes factors for effective teamwork that we will use to analyze the impact of community smells on software development teams' performance.

\subsection{Social debt and community smells}

Social debt is a concept coined by sociological researchers over 50 years ago~\cite{MuirEugine_SocialDebt:1962}.
This first definition of social debt adopted economic concepts to explain how the social obligations involved in repaying exchanged favors stress the social relationships among the people involved in the transaction, i.e., creditors and debtors.
The higher the number of strained relationships, the larger the social debt.

Building on this concept, software engineering researchers have adopted and applied social debt concepts to address the same phenomenon in software development teams and their organizations~\cite{Tamburri-etal_whatSDinSE:2013, Tamburri-etal_socialDebtSEinsights:2015}.
Over time, researchers have focused on measuring social debt and identifying aspects of software development teams that relate to the sources of social debt, i.e., community smells~\cite{Palomba-etal_beyondTechnicalAspects:2018,Tamburri-etal_exploringCSsOS:2019, Catolino-etal_genderDiversityWomen:2019,Tamburri_DAHLIA:2019}.
In other words, the presence of community smells suggests that something is going wrong within the software development teams, e.g., there are conflicts among teammates or there are no standards for the team to follow.
Over time, the presence of these community smells leads to the accumulation of social debt.

\subsection{Teamwork and factors for its effectiveness}

Reviews on human resource management and psychology provides a concise set of nine critical teamwork factors~\cite{Salas-etal_teamConsiderations:2015}.
These nine factors were prevalent in studies about teamwork and showed high impact on team results.
The factors include six team properties and three conditions that affect the team properties.
Table~\ref{table:table_9criticalFactorsTeamwork} shows the nine factors.

Because effective teamwork is necessary for team performance, we believe these factors can help produce additional valuable insight into the causes and effects of the community smells.
By analyzing the community smells in light of these factors, we can provide more details about how the community smells impact the intangible structure of teamwork and then the performance of software development teams.

\begin{table}[!ht]
\caption{Critical factors for teamwork effectiveness~\cite{Salas-etal_teamConsiderations:2015, Bell-etal_teamCompositionABCsteamwork:2018}}
\centering
\begin{threeparttable}
\begin{tabular}{ p{0.075cm} p{2.7cm} p{9.5cm}}
 \hline
 \rowcolor{Gray}
 \textbf{\#} & \textbf{Factor} & \textbf{Description} \\  
 \hline
 1 & Cooperation & The mix of motivational drivers necessary for teamwork: the team's attitudes, beliefs, and feelings.\\
 \hline
 2 & Conflict & The perceived incompatibilities in the interests, beliefs, or views held by at least one teammate.\\
 \hline
 3 & Coordination & The enactment of behavioral and cognitive mechanisms necessary to perform a task and transform team resources into outcomes.\\
 \hline
 4 & Communication & The reciprocal process of sending and receiving information that forms and reforms teams' attitudes, behaviors, and cognition.\\
 \hline
 5 & Coaching & The enactment of leadership behaviors to establish goals and set directions, leading to the successful accomplishment of these goals.\\
 \hline
 6 & Cognition & The shared understanding among teammates resulting from their interactions. Cognition includes knowledge of roles and responsibilities, team objectives and norms, and teammate knowledge, skills, and abilities.\\
 \hline
 7 & Composition\tnote{*} & The individual factors relevant to team performance: good teammate's features, the best configuration of teammate knowledge, skills, attitudes, and diversity.\\
 \hline
 8 & Context\tnote{*} & The situational characteristics or events that influence the occurrence and meaning of behavior, and the manner and degree to which various factors impact team outcomes.\\
 \hline
 9 & Culture\tnote{*} & The assumptions people hold about relationships with each other and the environment shared among a group of people and manifest in individuals' values, beliefs, norms for social behavior, and artifacts.\\
 \hline 
\end{tabular} 
\begin{tablenotes}\footnotesize
\item[*] Influencing conditions
\end{tablenotes}
\end{threeparttable}
\label{table:table_9criticalFactorsTeamwork}
\end{table}

\section{Methodology}
    \label{sec:3-3}
    We designed the methodology for this SLR following Kitchenham's approach~\cite{kitchenham_procedures:2004}.

\subsection{Defining the research questions}

Cunningham first described TD as the postponed implementation of software architecture decisions to reduce workload and allow for the release of the most suitable functional software products for customers in the shortest time~\cite{Cunningham_technicalDebtDefinition:1992}.
In other words, teams trade software quality to speed up the development process.
To repay the TD, teams must refactor the software.
Subsequent TD researchers have focused on offering more definitions~\cite{McConnell_definitionTD:2021, Kruchten_TDfromMetaphorPractice:2012, Shull_TDTransferEmpiricalResults:2013}, recognized other types of technical smells besides code smells~\cite{SHARMA_softwareSmells:2018,Alves_TDontology:2014, Izurieta_TDlandscape:2012}, causes~\cite{SHARMA_softwareSmells:2018}, and awareness~\cite{Ernst-etal_technicalDebtMeasureManage:2015}.
Other researchers have developed frameworks for managing TD in large software development projects~\cite{Guo_technicalDebtExploringCosts:2016}, agile software development projects~\cite{Behutiye-etal_analyzingTD:2017}, and startups~\cite{Giardino-etal_technicalDebtStartupCos:2016}.

While TD focuses primarily on the technical aspects of software, development teams can also face problems related to the social dimension of software projects.
A more recent concept in software engineering is \textit{social debt}~\cite{Tamburri-etal_whatSDinSE:2013}.
Researchers adopted the idea of social debt from sociological research~\cite{MuirEugine_SocialDebt:1962}.
TD and social debt have some similarities: risky decisions are common triggers that affect software developers and software quality.
However, there are differences.
TD focuses on postponed technical decisions and potential software failures~\cite{SHARMA_softwareSmells:2018}.
Social debt focuses on wrong socio-technical decisions that shape work environments.
These decisions then influence the  welfare, social interactions, and performance of software developers~\cite{Tamburri-etal_socialDebtSEinsights:2015}.
So, while TD focuses on the technical aspects of software, social debt focuses on people.

As an initial step in clearly understanding social debt in software engineering and the differences between social debt and TD, we start by analyzing definitions.
Therefore, the first research question is:

\vspace{8pt}
\begin{tcolorbox}
\textbf{RQ1 - How have software engineering researchers defined \textit{social debt}?}
\end{tcolorbox}
\vspace{8pt}

Social debt accumulation can start with one or more wrong socio-technical decisions.
The more bad decisions, the greater the negative effects.
As long as the bad decisions remain unfixed, their effects persist and their intensity increases.
As an analogy to code smells in TD, the connection between wrong socio-technical decisions and adverse effects on software organizations are called \textit{community smells}.
The ongoing presence of community smells leads to the accumulation of social debt.

The community smells play a critical role in disrupting the organizational and social structures that define software engineering communities.
Besides social interactions, community smells impact organizational finances and reputation due to low-quality software products affecting customers.
This chain of events can put the business continuity of software development organizations at risk.
To provide a broad overview of the potential sources of social debt, the second research question is:

\vspace{8pt}
\begin{tcolorbox}
\textbf{RQ2 - What community smells appear in the literature?}
\end{tcolorbox}
\vspace{8pt}

The prevalence of community smells in software engineering communities lead to social debt.
Such a dynamic process can be from social, technical, and organizational perspectives.
Consequently, it is crucial to identify and examine how management approaches mitigate the community smell effects and pay back the social debt.
Therefore, the third research question is:

\vspace{8pt}
\begin{tcolorbox}
\textbf{RQ3 - What approaches for community smells management appear in the literature?}
\end{tcolorbox}
\vspace{8pt}

Social debt is a relatively young topic in software engineering.
The social debt literature does include studies that examine the effects of community smells in the software development process.
However, it is difficult to obtain a consistent understanding of the characteristics of community smells because the information is scattered throughout the available studies.
Therefore, it is necessary to construct a list of properties along with their descriptions.
Based on these motivations, the fourth research question is:

\vspace{8pt}
\begin{tcolorbox}
\textbf{RQ4 - What are the properties of the community smells?}
\end{tcolorbox}
\vspace{8pt}

A previous paper indicated the need for more research to better understand the birth and growth of community smells~\cite{Tamburri_splicingCSASmells:2019}.
To address this need, the fifth research question is:

\vspace{8pt}
\begin{tcolorbox}
\textbf{RQ5 - How do community smells originate and evolve in software development settings?}
\end{tcolorbox}
\vspace{8pt}

The standard representation for community smells is through cause and effect models.
However, we observed that the causes of the community smells described events occurring or emerging in different ways.
Prior to our formal analysis of the papers, we identified the need to characterize the occurrence of causes and effects of community smells using probability theory about types of events, i.e., dependent or independent.
This analysis can offer valuable insights into the patterns that characterize the occurrence of the causes and effects of community smells.
Therefore, the sixth research question is:

\vspace{8pt}
\noindent
\begin{tcolorbox}
\textbf{RQ6 - What types of events characterize the causes and effects of community smells?}
\end{tcolorbox}
\vspace{8pt}

Prior research described an approach to categorize community smells based on teamwork factors, i.e., cooperation, communication, and coordination~\cite{Tamburri-etal_socialDebtSEinsights:2015, Tamburri-etal_exploringCSsOS:2019}.
However, the small number of factors and the limited number of community smells that researchers have characterized motivated us to search for robust theories on teamwork factors to extend the prior work.
In addition, there is a need to characterize all of the causes and effects of community smells and establish their connections with bad teamwork practices.
These connections can offer more insight into poor performance in software development teams.
Therefore, the last two research questions are:

\vspace{8pt}
\noindent
\begin{tcolorbox}
\textbf{RQ7 - What are the types of causes and effects of community smells found in the literature?}
\end{tcolorbox}
\vspace{8pt}

\vspace{8pt}
\noindent
\begin{tcolorbox}
\textbf{RQ8 - How do the community smells affect teamwork in software development teams?}
\end{tcolorbox}
\vspace{8pt}

\subsection{The source selection and search}

To identify more relevant studies, we used three keywords.
First, we included the keyword \textit{community smell} because it is the focus of our work.
Second, community smells are part of the larger \textit{social debt} concept.
Third, we included \textit{software engineering} to help us exclude studies from other domains, like social science where social debt originated~\cite{MuirEugine_SocialDebt:1962}.
With the aim of guaranteeing the coverage of research on community smells as sources of social debt in software engineering, we executed query strings on the advanced search interfaces of the following databases: \textit{ACM}, \textit{IEEE Xplore}, \textit{Science Direct}, \textit{Springer}, \textit{Scopus}, and \textit{Science Citation Index (Web of Science)}, and \textit{Google Scholar}.

To ensure broad coverage and identify papers that describe community smells in various research contexts, we searched in \textit{full text}, \textit{all fields}, and \textit{anywhere}.
Table~\ref{table:table_searchStrings} provides the search details for each database.

\begin{table}[!ht]
\caption{Search Strings}
\centering
\begin{tabular}{ p{0.1cm} p{1.75cm} p{6.4cm} p{3.8cm}}
 \hline
 \rowcolor{Gray}
 \textbf{\#} & \textbf{Database} & \textbf{Search string} & \textbf{Search based on}\\
 \hline 
1 & ACM DL & ("community smells" AND "social debt") AND "software engineering" & Keywords anywhere\\
 \hline
2 & IEEE Xplore & ("community smells" AND "social debt") AND "software engineering" & Keywords in full text and metadata\\ 
 \hline
3 & Science Direct & ("community smell" OR "social debt") AND "software engineering" & Keywords\\
 \hline
4 & Springer & social AND debt AND software AND engineering AND "community smells" & Words and an exact phrase\\
 \hline
5 & SCOPUS & ("community smells" AND "social debt") AND "software engineering" & Keywords in all fields\\
 \hline
6 & Web of Science & TS=(community smells* OR social debt) AND SU=Computer Science & Topic \& research area\\
 \hline
7 & Google Scholar & "community smells" + "social debt" + "software engineering" & Keywords anywhere\\ 
 \hline 
\end{tabular} 
\label{table:table_searchStrings}
\end{table}

\subsection{Selection criteria}

Table~\ref{table:table_inclusionExclusionCriteria} provides the inclusion/exclusion criteria.
While most of the selection criteria are standard, we did need some specific inclusion criteria because community smells and social debt are relatively new topics in software engineering.
The first criterion helped us select those papers examining community smells as socio-technical triggers of social debt.
Although an SLR regularly compiles empirical studies, our second inclusion criterion helps identify non-empirical studies to cover the development of new theories and corresponding future work.
Also, the sixth inclusion criterion is there because the seminal paper on social debt was published in 2013~\cite{Tamburri-etal_whatSDinSE:2013}.

\begin{table}[!ht]
\caption{Inclusion and exclusion criteria}
\centering
\begin{tabular}{ p{0.1cm} p{6.3cm} p{0.1cm} p{5.5cm} }
 \hline
 \rowcolor{Gray}
 \textbf{\#} & \textbf{Inclusion criteria} & \textbf{\#} & \textbf{Exclusion criteria} \\
 \hline
  1 & Relevant studies to community smells as phenomena explaining social debt accumulation in software engineering & 1 & Studies not written in English\\
  2 & Empirical and theoretical studies, either qualitative or quantitative, examining community smells & 2 & Unavailable or unpublished studies\\
  3 & Studies discussing approaches to manage community smells & 3 & Extended abstracts\\
  4 & Replication studies & 4 & Position and duplicated papers\\
  5 & Publications from peer-reviewed venues such as journals, conferences, and workshops & 5 & Book chapters\\
  6 & Studies published since 2013 & 6 & Tutorials and magazines\\
 \hline 
\end{tabular}

\label{table:table_inclusionExclusionCriteria}
\end{table}

\subsection{Review execution}

The search identified 95 studies.
The first author read the abstracts, keywords, and conclusions for each study and determined that the information in these sections was insufficient for determining inclusion and exclusion.
Thus, the first applied the inclusion and exclusion criteria to examine the full text of the papers.
This full-text review identified in 25 studies for inclusion.

To illustrate the reason for this step, the paper ``Interpersonal conflicts during code review''~\cite{Wurzel_etal_InterpersonalConflictsCodeReview:2022} was not relevant according to the abstract, keywords, and conclusion.
However, a full-text review of the paper, searching for ``community smells,'' shows that it is relevant because it discusses the relationship between interpersonal conflicts and community smells.

As another illustration, a study about challenges to DevOps in industry~\cite{Caprarelli_etal_FallaciesPitfallsDevOps:2020} did not contain explicit information about community smells in the abstract, keywords, and conclusion.
However, a full-text analysis found the study discussed eight challenges that undermine the adoption of DevOps practices.
Three of those eight challenges reflected the effects of community smells, where a limited number of qualified personnel was a constant source of problems.
Consequently, process delay or waste of time was a common community smell effect.

To validate the first author's results, the second author applied the inclusion and exclusion criteria to a randomly selected 20\% of the studies.
The first and second authors agreed on all of these papers.
Then the first author went back and examined the remaining papers to ensure there were no potential problems.
This review resulted in no other changes.

Finally, we used a standard backward snowballing process.
After identifying the final list of included studies from the literature search above, we checked the references of each study, using the same inclusion approach described above.
However, we did not identify any additional papers that needed to be included.

\subsection{Data extraction}

To extract data about community smells from the 25 included studies, we used the data items shown in Table~\ref{table:table_itemsDataExtraction}.
This table also indicates which research questions motivated the inclusion of each data item.

\begin{table}[!ht]
\caption{Items for data extraction}
\centering
\begin{tabular}{ p{2.8cm} p{8.5cm} p{1.1cm}}
 \hline
 \rowcolor{Gray}
 \textbf{Data item} & \textbf{Description} & \textbf{RQ}\\ 
 \hline 
Study identifier & A unique digit identifier for a paper & -\\
 Title & The title of the paper & -\\
 Research contributions & Research results, e.g., definitions, models, and tools & 1, 2, 3\\
 Property & A qualitative or quantitative attribute describing the nature of community smells. & 4, 5, 6 \\
 Community smell & Because the representation of community smells is a causality model, this data item registers the names assigned to every model. & 7, 8\\
 Causes & The sociotechnical events triggering the effects of community smells. & 7, 8\\
 Effects & The sociotechnical events produced by the causes of community smells. & 7, 8\\
 Type of cause & The feature assigned to a community smell cause based on its connection with a critical teamwork factor. & 7, 8\\
 Type of effect & The feature assigned to a community smell effect based on its connection with a critical teamwork factor. & 7, 8\\
 \hline 
\end{tabular}
\label{table:table_itemsDataExtraction}
\end{table}

\subsection{Data analysis}

To answer RQ1, we first listed the definitions of social debt identified in the literature.
Next, we include definitions of two related concepts: \textit{socio-technical debt} focuses on problems that arise from social causes; and \textit{social sustainability debt} focuses on the impact social debt has on society.
Finally, we synthesized this information to provide a common understanding of social debt in software engineering.

For RQ2, we extracted information that describes the origin and causal relationship of each community smell.
The information included the causes, effects, and descriptions of each smell.
While the literature did indicate the possibility of additional community smells, the ones presented here are the ones that were most frequently reported.

To answer RQ3, we collected data on the research contributions from the included studies.
The data included the contribution name, description, and application.
Then we performed a \textit{topic-independent classification} according to Petersen, et al.'s updated guidelines~\cite{PETERSEN_systematicMappingUpdate:2015}.
We listed and explained categories to classify the types of research contributions according to the context of our study.
\ref{appendix_A} provides the details for each category.

To answer RQ4, the first author extracted data from the results and discussions of the included studies to construct the features and behaviors of community smells.
Then he drafted a list of properties and their descriptions for the community smells.
The second author reviewed the properties and suggested adjustments to arrive at the final version.

For RQ4, to provide a straightforward explanation of the origin and evolution of community smells, we borrowed a metaphor from public health to represent the community smells as chronic conditions.
A chronic condition is ``\textit{a health condition or disease that is persistent or otherwise long-lasting in its effects or a disease that comes with time}''~\cite{Bernell_etall_whatisChronicDisease:2016}.
Thus, it seemed fitting to use this metaphor to describe community smells that behave like chronic conditions in software development teams and organizations.

We used the National Cancer Institute's framework for the stages of cancer (a chronic condition)\footnote{\href{https://www.cancer.gov/about-cancer/diagnosis-staging/staging}{Cancer staging systems}} to represent the origin and evolution of community smells.
This stage-based representation can help teams diagnose the status of community smells as their effects worsen and spread.
The first author developed the initial version of the framework.
The second author reviewed and suggested adjustments to arrive at the final version.

To answer RQ6, the first author extracted the causes and effects of the community smells.
Then all three authors met to review the collected data.
The authors recognized that causes had different patterns in how they occur or emerge.
Therefore, the authors agreed to characterize the causes and effects of the 30 community smells.
The first author characterized the causes and effects as either \textit{dependent}, meaning they rely upon other events or \textit{independent}, meaning they do not.
Finally, the second author reviewed the first author's work and made suggestions to arrive at the final version.

To answer RQ7 and RQ8, we coded information about the causes and effects of the community smells.
The following subsections explain the design and execution of our coding process.

\subsubsection{Defining the need for coding}

The three authors met and reviewed the causes and effects of the 30 community smells.
Because we observed large amount of qualitative information present, we decided it was important to code this information to produce more concrete results.

\subsubsection{Generating the codes}
The first step in the coding process is to generate the set of codes.
To generate the codes, the first author reviewed teamwork literature~\cite{Salas-etal_teamConsiderations:2015, Bell-etal_teamCompositionABCsteamwork:2018} and the papers included in the SLR.
He extracted high-level concepts, sub-concepts, and descriptions.
Then he organized the concepts into \textit{core categories} (based on high-level concepts) and \textit{subcategories} (based on sub-concepts).
The 12 core categories consisted of the nine critical factors for effective teamwork (Table~\ref{table:table_9criticalFactorsTeamwork}) and three concepts from the included studies: economic impact, technical debt effect, and community smell.
The 29 sub-concepts comprised the subcategories.
\ref{appendix_B} provides the list of codes and descriptions.

\subsubsection{Reviewing the codes}

The three authors met and reviewed the initial list of codes and descriptions.
The second and third authors made suggestions to clarify the codes.
The three authors then met again and tested out the codes on a random set of causes and effects for the community smells.
The results of this test indicated that the codes were useful for characterizing the sample of causes and effects.
Therefore, we launched the process of coding the full dataset.

\subsubsection{Conducting the coding}

We used an iterative coding process, conducted primarily by the first and third authors.
For the initial iteration, the first author performed the high-level and fine-grained coding.
Using a online spreadsheet as an interaction mechanism, he provided the code along with his rationale.
Then, the third author went through these comments and either agreed or provided feedback about how to adjust the codes.

In cases where the third author did not fully understand the code or the rationale, the authors discussed the cause or effect to arrive at a common understanding.
For instance, the authors discussed some of the details about the causes and effects of community smells described in the papers to ensure we had the same understanding.
In other cases, we had to infer information from the papers when the information was either ambiguous or incomplete.

In addition, the structure of the community smells consists of causes and effects.
Most causes and effects include two or more specific socio-technical events.
We split the descriptions of causes and effects containing multiple socio-technical events into more atomic items.
However, some causes and effects still required two codes to characterize them.
We used the label dual nature for double codes as we wanted to identify the essence of every single cause and effect of community smells.

Over the whole process, we interacted through 333 comments in the online spreadsheets.
After this process, all three authors met to discuss the strategies to visualize the results.

\subsection{Visualization}

We used \textit{R} to plot the results on bar charts and Sankey diagrams.
A Sankey diagram is a flow diagram used to illustrate the flow of values (e.g. energy, material, or cost) within systems~\cite{Schmidt_SankeyEnergyMaterialFlow:2008, BAKENNE_sankeyHydrogenEconomy:2016}.
The Sankey diagrams have been useful in supporting the analysis of systems flows in medicine~\cite{ICAY_SankeyMolecularData:2018} and energy~\cite{SOUNDARARAJAN_sankeyFrameworkforEnergy:2014,SUBRAMANYAM_sankeyMapEnergyFlow:2015,BAKENNE_sankeyHydrogenEconomy:2016}.
These studies show that Sankey diagrams help clearly visualize the connections among multiple variables.
Because our work links causes to community smells to effects, we found Sankey diagrams to be a helpful visualization of the relationships.

In our case, the Sankey diagrams consist of three sets of nodes to show the connections between the causes and effects of community smells.
The nodes on the left side of the diagrams represent the causes of community smells.
The nodes in the middle represent the community smells.
The node on the right side of the diagram represents the effects, categorized by teamwork factors (e.g., cooperation, coordination, and communication).

The flows in the diagrams represent connections between the causes and effects for the community smells and teamwork factors.
The label for each node also includes the number of incoming and outgoing connections.
The relationships represented here are many-to-many relationships.

\section{Reporting the results}
    \ref{appendix_C} lists the 25 studies included in the SLR.
We categorized 24 studies as empirical and one as theoretical according to \ref{appendix_A}.
Empirical studies conducted by Tamburri~\cite{Tamburri-etal_socialDebtSEinsights:2015,Tamburri-etal_exploringCSsOS:2019,Tamburri_DAHLIA:2019, Tamburri-etal_architectsRole:2016} and Palomba~\cite{Palomba-etal_beyondTechnicalAspects:2018} offer information about the identification of the causes and effects of community smells in practice, information about the impact on work settings, and management approaches.
The remaining included studies provide further information on community smells in relationship with development communities and processes~\cite{DeStefano_etal_SplicingCommunityPatternsSmells:2020,Tamburri_splicingCSASmells:2019, Wurzel_etal_InterpersonalConflictsCodeReview:2022,   Caprarelli_etal_FallaciesPitfallsDevOps:2020, DeStefano_etal_ImpactsSoftwareCommunityPatternsProcessProduct:2022,  Tambuerri_etal_OmniscientDevOpsAnalytics:2019, Brenner_11NontechnicalPhenomenaTD:2019, Lavallee_multiTeamsImpactSQ:2018} and research contributions such as models~\cite{Catolino-etal_genderDiversityWomen:2019,Palomba_etal_UnderstandingCSsVariability:2021,Catolino_etal_GenderDiversityCommunitySmellsInsightsTrenches:2019, Eken_etal_EmpiricalStudyEffectCommunitySmellsBugPrediction:2021, Almarimi_etal_LearningDetectCommunitySmellsOSSP:2020, Almarimi_etal_DetectionCommunitySmellsGPEClassifierChain:2020,Almarimi_etal_CsDetector:2021, Huang_etal_PredictingCommunitySmellsOccurrenceSentiments:2021, Huang_etal_CommunitySmellOccurrencePredictionMulti-Granularity:2022,Palomba_Tamburri_PredictingEmergenceCommunitySmellSTMetrics:2021}, organizational strategies~\cite{Catolino_etal_RefactoringCommunitySmells:2020}, and tools~\cite{Tamburri-etal_discoveringCommunityPatterns:2019}.
We organized our results around the research questions described in Section~\ref{sec:3-3}.

\subsection{RQ1: How have software engineering researchers defined \textit{social debt}?}

The first researchers to define social debt in software engineering based their definition on sociological concepts that help explain how socio-technical problems affect software projects.
This seminal work informally defined social debt as ``\textit{unforeseen project costs connected to a suboptimal development community}.''
The initial description of social debt focused on unforeseen project costs resulting from poor socio-technical decisions that affect the social interaction and performance of software development communities.
This work describes the primary difference between technical debt and social debt is that the latter impacts people~\cite{Tamburri-etal_whatSDinSE:2013}.

A later empirical study identified the core concepts that shaped the formal definition of social debt.
An interpretative framework modeled the interaction among these concepts and demonstrated how software development organizations could accumulate social debt.
The elements that play a crucial role in explaining social debt include: \textit{Socio-technical decisions}, \textit{community smells}, \textit{sub-optimal team organizational}, and \textit{sub-optimal social structures} .
Thus, the authors formally define social debt as ``\textit{a cumulative and increasing cost in the current state of things, connected to invisible and adverse effects within a development community}''~\cite{Tamburri-etal_socialDebtSEinsights:2015}.

The relationship between social debt and technical debt suggests another important concept.
\textit{Socio-technical debt} is the result of social debt that includes technical problems in software artifacts, e.g., documentation and software components.
Like social problems, these technical problems can also arise from incorrect socio-technical decisions.
In other words, TD can be a consequence of social debt~\cite{Tamburri_DAHLIA:2019}.

Finally, \textit{social sustainability debt} is \textit{the hidden effect of past decisions about software systems that negatively affect social justice, equity, and fairness, or which lead to an erosion of trust in society}~\cite{Betz-etal_sustainabilityDebt:2015}.
For instance, software built based on wrong socio-technical decisions that lead to social debt may affect social relationships between the developers and society, e.g. customer organizations or social communities.
Because later papers include aspects of social sustainability debt as part of the standard definition of social debt (e.g. by including community smells related to software development organizations or society~\cite{Tamburri-etal_architectsRole:2016}), we do not find any concepts that expand the definition of social debt.

However, social sustainability debt may be a helpful category for community smell effects that impact the final software users in society.

In summary, organizations that incur social debt have leaders or project managers who make decisions that negatively impact people's well-being, team performance, and software quality.
When these wrong decisions remain unfixed, software development teams and organizations accumulate socio-technical effects, including  1) adverse human reactions, 2) strained social relationships, 3) faulty software artifacts, 4) low-quality software products, and 5) and economic impact.
Therefore, organizations need to address poor socio-technical decisions and repay the social debt.

\subsection{RQ2: What community smells appear in the literature?} \label{subsec:Ch2RQ2}

The 30 community smells listed and described in Table~\ref{longtable:table_communitySmellsDescriptions} are those most frequently reported in the software engineering literature~\cite{Tamburri-etal_socialDebtSEinsights:2015,Tamburri_DAHLIA:2019, Tamburri-etal_architectsRole:2016,  Palomba-etal_beyondTechnicalAspects:2018, Tamburri-etal_exploringCSsOS:2019}.
Researchers represent a community smells through a cause and effect model.
\ref{appendix_D} shows the causes and effects of the 30 identified community smells.
The \textit{causes} are adverse social and organizational situations resulting from the incorrect socio-technical decisions.
The \textit{effects} depict the impact of the incorrect socio-technical decisions on software development teams, e.g., tense social interactions, personal and professional misconduct, and faulty software artifacts.

\begin{small}
\begin{longtable}[p]{p{0.02\textwidth} p{.2\textwidth} p{.68\textwidth}}
\caption{Community smells found in the literature\label{long_2p0}}\\
 \hline
	\textbf{\#} & \textbf{Name} & \textbf{Description} \\ \hline
 \endfirsthead
\hline
 \multicolumn{3}{c}
 {{\bfseries \tablename\ \thetable{} -- continued from previous page}} \\
\hline
\textbf{\#} & \textbf{Name} & \textbf{Description}\\
\hline 
 \endhead
\hline
 \multicolumn{3}{r}{{Continued on next page}} \\ \hline
\endfoot
\hline
\endlastfoot

1 & Architecture by osmosis & This smell describes people making architecture decisions based on inappropriate information management. While some teammates manage and filter information through undefined communication channels or methods, others use different standards to document change requests. These actions cause problems with locating architecture decision sources, waste of time, and unstable architecture configuration~\cite{Tamburri_DAHLIA:2019}.\\ \hline

2 & Architecture hood & The smell emerges when decision-making teams in charge of architecture decisions work remotely or distant from teammates. Thus, cooperation is challenging. Developers cannot exchange communications with the software architects responsible for those decisions causing implementation problems~\cite{Tamburri-etal_socialDebtSEinsights:2015}.\\ \hline

3 & Black cloud & This smell is present when organizations do not provide the conditions for social interactions and effective communication between teammates. Thus, the conditions do not support the exchange of knowledge during software development processes, e.g., professional experience or understanding of projects in progress~\cite{Tamburri-etal_socialDebtSEinsights:2015, Palomba-etal_beyondTechnicalAspects:2018}.\\ \hline

4 & Class cognition &  In this smell, refactoring makes the modular structures and refactored classes more difficult to understand. Thus, other teammates, including current developers and newcomers, spend time and effort understanding the new environment version and file locations~\cite{Palomba-etal_beyondTechnicalAspects:2018}.\\ \hline

5 & Code red & It is a smell denoting the existence of highly complex classes. Thus, a limited number of developers, i.e., 1–2 people at most, can manage such classes~\cite{Palomba-etal_beyondTechnicalAspects:2018}.\\ \hline

6 & Cognitive distance & In software engineering, it is the distance that developers perceive on the physical, technical, social, and cultural levels concerning peers with considerable background differences. The effects are associated with, for example, resource management issues, conflicts among teammates, and code smells~\cite{Tamburri-etal_architectsRole:2016}.\\ \hline

7 & Cookbook development & Developers are stuck in how they usually work according to old-framework-based cookbooks, e.g., the waterfall model. They do not accept innovative ideas or new ways of working, e.g., agile methods, that change their comfort zones. Consequently, software product features do not fulfill customer’s expectations~\cite{Tamburri-etal_architectsRole:2016}.\\ \hline

8 & DevOps clash & The smell describes clashes between development and operations teams from multiple geographical locations with contractual obligations to either development or operations activities. These clashes lead to a slower development process and ineffective operations~\cite{Tamburri-etal_architectsRole:2016}.\\ \hline

9 & Disengagement & It is a situation where developers think the software product is mature enough. Then they send it to operations technicians even though the software is unfinished. This circumstance reflects a lack of curiosity from the developers' side, ending in missing software features or applying wild assumptions~\cite{Tamburri-etal_architectsRole:2016}.\\ \hline

10 & Dispersion & This community smell concerns a fix or refactoring that causes the fragmentation of an existing group, working in or being part of a collaboration network. Functionality rearrangements also lead to haphazard work. Finally, coordinating and carrying out maintenance activities is more challenging~\cite{Palomba-etal_beyondTechnicalAspects:2018}.\\ \hline

11 & Dissensus & It is the inability to achieve consensus on how to proceed despite repeated attempts at doing it. Therefore, code smells detected in software components remain without the required adjustments~\cite{Palomba-etal_beyondTechnicalAspects:2018}.\\ \hline

12 & Hyper community & The smell refers to a highly connected community sensible to groupthink. It also influences smaller subcommunities in its network. Consequently, it leads to social turbulence and faulty software components~\cite{Tamburri-etal_architectsRole:2016}.\\ \hline

13 & Informality excess & Excessive informality is the relative absence of information management and control protocols. These conditions lead to information spillover and low accountability of teammates~\cite{Tamburri-etal_architectsRole:2016}.\\ \hline

14 & Institutional isomorphism & It is the similarity of processes or structures of one software development group or sub-community to those of another. This condition can be the result of imitation or independent development under the same constraints. The effects can include lack of innovation, stagnancy, and communication~\cite{Tamburri-etal_architectsRole:2016}.\\ \hline

15 & Invisible architecting & It is a situation in which teammates document software architecture decisions and register meeting agreements inconsistently. Consequently, the descriptions of such decisions, necessary for the software architecture process, become invisible. The lack of accurate information generates, for example, decision unawareness and conflicts among teammates~\cite{Tamburri_DAHLIA:2019}.\\ \hline

16 & Leftover techie & This smell consists of a broken collaborative network. Also, there is no effective communication between developers and technicians (e.g., help desk, operation, and maintenance). The technicians are not usually involved in multiple aspects of the software development process to have a shared knowledge of software features~\cite{Tamburri-etal_socialDebtSEinsights:2015}.\\ \hline

17 & Lone wolf & This smell occurs when defiant teammates carry out their work irrespective or regardless of their peers. This smell reflects poor communication addressing project needs. The effects are, for instance, unsanctioned architecture decisions across the development process, code smells, and project delays~\cite{Tamburri-etal_exploringCSsOS:2019, Catolino-etal_genderDiversityWomen:2019}.\\ \hline

18 & Lonesome architecting & Non-architect teammates see the need to make architecture decisions because the current architects are too few and far apart. These non-architects make decisions without consulting with experts involved in such decisions. From a social perspective, developers are unaware of what they are doing. Also, this scenario leads to incompatibility problems and faster decision-making~\cite{Tamburri_DAHLIA:2019}.\\ \hline

19 & Newbie free-riding & Newcomer teammates must understand by themselves what to do and for whom, which leads to the free-riding of older employees. This unfriendly work environment adds more pressure to the team and affects people's psychological state~\cite{Tamburri-etal_architectsRole:2016}.\\ \hline

20 & Obfuscated architecting & It occurs when multiple subteams of a development network lack the organizational and the socio-technical vision required for harmonized operations. For instance, projects need new developers to implement changes to legacy and new products. However, newcomers included in projects lack the technical background to deal with legacy products. Among effects, teams faced waste of time, code smells, and developers' frustration~\cite{Tamburri_DAHLIA:2019}.\\ \hline

21 & Organizational silo & This community smell is associated with task coordination problems. Software development tasks are sometimes not interconnected with each other. This smell sets challenges to check task dependencies and non-conducive conditions for effective communication among teammates performing tasks. This scenario puts the socio-technical congruence at risk~\cite{Tamburri-etal_socialDebtSEinsights:2015, Tamburri-etal_exploringCSsOS:2019}.\\ \hline

22 & Organizational skirmish & An organizational skirmish is a scenario where teams have differences over their organizational cultures. It makes the work of project managers difficult. The impact is notorious on productivity, e.g., project delays~\cite{Tamburri-etal_socialDebtSEinsights:2015}.\\ \hline

23 & Power distance & The distance that less powerful software development teammates perceive with those power-holder teammates, e.g., experienced teammates or decision-makers. It finally disrupts the software process and impacts organizational finances~\cite{Tamburri-etal_architectsRole:2016}.\\ \hline

24 & Priggish members & They are pedant teammates demanding of others pointlessly precise conformity or exaggerated propriety, which is annoying. The attitude frustrates teammates and affects the software process~\cite{Tamburri-etal_architectsRole:2016}.\\ \hline

25 & Prima donnas & This smell indicates the presence of teammates working in isolation. They are unwilling to welcome the change of legacy products and support from other teammates. These teammates prevent the organization from innovative solutions or processes and effective communication and collaboration~\cite{Tamburri-etal_socialDebtSEinsights:2015}.\\ \hline

26 & Radio silence or Bottleneck & It is a scenario where leaders and teammates perform tasks in very formal and complex organizations. Under these conditions, team communication structures are not conducive to spread information across the teams efficiently. For instance, a person working as a unique information intermediary for different teams leads to communication overload and massive delays~\cite{Tamburri-etal_socialDebtSEinsights:2015, Tamburri-etal_exploringCSsOS:2019}.\\ \hline

27 & Sharing villainy & This community smell depicts work environments where the goal of sharing reliable knowledge or information is a challenge. Organizations do not provide the expected conditions for knowledge sharing, like opportunities for meetings and incentives. Hence, some teammates do not see knowledge sharing as a productive activity for projects' completion~\cite{Tamburri-etal_socialDebtSEinsights:2015}.\\ \hline

28 & Solution defiance & This smell describes conflicts between teams. Teammates with similar deep-level factors (e.g., technical background) and organizational cultural beliefs (e.g., values and norms) create subteams. Then they go into conflicts in decision-making meetings since every group supports their opinions on potential solutions~\cite{Tamburri-etal_socialDebtSEinsights:2015}.\\ \hline

29 & Time warp & After changes in organizational structures and processes, experienced teammates wrongly assume time frames for exchanging communications and no need for explicit coordination. Some effects of this community smell are unsatisfied customers and faulty software~\cite{Tamburri-etal_architectsRole:2016}.\\ \hline

30 & Unlearning & Components of training courses, e.g., technological innovations and best practices, become unfeasible learning material when it is shared with older teammates. These teammates show a very high experience diversity. Consequently, the updated accumulated knowledge is at risk of being gradually lost~\cite{Tamburri-etal_architectsRole:2016}.
\label{longtable:table_communitySmellsDescriptions}
\end{longtable}
\end{small}

\subsection{RQ3: What approaches for community smells management appear in the literature?}
\label{mitigations}

This section describes the five types of management approaches we identified based on the classification scheme in \ref{appendix_A}: organizational strategies, frameworks, models, tools, and guidelines.
The management approaches can help monitor and mitigate community smells and pay back the accumulated social debt.
Table~\ref{longtable:table_managementApproaches} provides an overview of the approaches.
The following subsections provide more detail on each type.

\begin{small}
	\begin{longtable}[p]{p{.18\textwidth} p{.3\textwidth} p{.42\textwidth}}
		\caption{Approaches to managing social debt\label{long_2p1}}\\
		\hline
		\textbf{Group} & \textbf{Approach} & \textbf{Key aspects}\\ 
\hline
 \endfirsthead
\hline
 \multicolumn{3}{c}
 {{\bfseries \tablename\ \thetable{} -- continued from previous page}} \\
\hline
\textbf{Group} & \textbf{Approach} & \textbf{Key aspects}\\ 
\hline
 \endhead
\hline
 \multicolumn{3}{r}{{Continued on next page}} \\ \hline
\endfoot
\hline
\endlastfoot
Organizational strategies &	Socio-technical decisions & Strategies aimed at mitigating the effects of community smells in practice~\cite{Tamburri-etal_socialDebtSEinsights:2015,Catolino_etal_RefactoringCommunitySmells:2020,Tamburri_DAHLIA:2019,Tamburri-etal_architectsRole:2016}\\
\hline
\multirow{11}{*}{Frameworks} & DAHLIA & Metrics-based framework: decision popularity, decision awareness, mean architecture incommunicability, and average per-person delay~\cite{Tamburri_DAHLIA:2019}\\
\cline{2-3}
& Socio-technical Quality framework & 40 socio-technical quality metrics comprising five categories: Developer Social network, Sociotechnical, Core Community Members, Turnover, and Social Network Analysis~\cite{Tamburri-etal_exploringCSsOS:2019, Palomba_etal_UnderstandingCSsVariability:2021, Palomba_Tamburri_PredictingEmergenceCommunitySmellSTMetrics:2021}\\
\hline
\multirow{18}{*}{Models} & Statistical & Prediction models for correlating community smells and code smells~\cite{Palomba-etal_beyondTechnicalAspects:2018}, predicting bug-prone code components~\cite{Eken_etal_EmpiricalStudyEffectCommunitySmellsBugPrediction:2021}, examining the impact of gender diversity in teams on mitigating community smells~\cite{Catolino-etal_genderDiversityWomen:2019, Catolino_etal_GenderDiversityCommunitySmellsInsightsTrenches:2019}, automatically detecting community smells in open-source projects with high precision~\cite{Almarimi_etal_LearningDetectCommunitySmellsOSSP:2020, Almarimi_etal_DetectionCommunitySmellsGPEClassifierChain:2020, Almarimi_etal_CsDetector:2021}, and predicting community smells on software developers based on sentiment analysis~\cite{Huang_etal_PredictingCommunitySmellsOccurrenceSentiments:2021, Huang_etal_CommunitySmellOccurrencePredictionMulti-Granularity:2022}\\ 
\cline{2-3}
& Social Network & Social-network-based models for building formal definitions of community smells~\cite{Tamburri-etal_exploringCSsOS:2019} and quantify forms of social debt in software architecture activities~\cite{Tamburri_DAHLIA:2019}\\
\hline
\multirow{7}{*}{Tools} & CodeFace4Smell & Extension of CodeFace tool that detects \textit{Organizational silo}, \textit{Black cloud}, \textit{Lone wolf}, and \textit{Radio silence}~\cite{Tamburri-etal_exploringCSsOS:2019}\\
\cline{2-3}
& YOSHI & Identify development communities, diagnose their health based on six factors, and potential for monitoring social debt~\cite{Tamburri-etal_discoveringCommunityPatterns:2019}\\
\hline
\multirow{3}{*}{Guidelines} & Managing team composition & List of 13 software engineering communities and their relevant features~\cite{Tamburri-etal_architectsRole:2016}
		\label{longtable:table_managementApproaches}
	\end{longtable}
\end{small}

\subsubsection{Organizational strategies}

Organizational strategies are socio-technical decisions designed and implemented by the organizations studied by researchers~\cite{Tamburri-etal_socialDebtSEinsights:2015, Tamburri-etal_architectsRole:2016, Tamburri_DAHLIA:2019}.
The purpose of these strategies is to mitigate community smells and pay back the accumulated social debt.
We identified 35 organizational strategies in the literature.
Table~\ref{longtable:table_organizationalStrategiesDescriptions} shows the organizational strategies, the target community smells, and the mitigation effectiveness.

\begin{small}
	\begin{longtable}[p]{p{.48\textwidth}  p{0.25\textwidth} p{.17\textwidth}}
		\caption{Organizational strategies found in the literature\label{long_2p2}}\\
		\hline
		\textbf{Organizational strategy} & \textbf{Community smell} & \textbf{Effectiveness} \\
\hline
 \endfirsthead
\hline
 \multicolumn{3}{c}
 {{\bfseries \tablename\ \thetable{} -- continued from previous page}} \\
\hline
 \textbf{Organizational strategy} & \textbf{Community smell} & \textbf{Effectiveness} \\
\hline 
 \endhead
\hline
 \multicolumn{3}{r}{{Continued on next page}} \\ \hline
\endfoot
\hline
\endlastfoot

Stand-up voting~\cite{Tamburri-etal_socialDebtSEinsights:2015} & Architecture hood & Partial \\
\hline

Create communication plan~\cite{Catolino_etal_RefactoringCommunitySmells:2020} & Black cloud & Full \\
\hline

\multirow{5}{*}{Social wiki~\cite{Tamburri-etal_socialDebtSEinsights:2015}} 
& Black cloud & \multirow{5}{*}{No data\footnotemark[1]} \\ 
& Organisational silo & \\
& Prima donnas & \\
& Sharing villainy & \\
& Solution defiance & \\
\hline

Architecture knowledge exchange through redocumentation, workshops, and presentations~\cite{Tamburri-etal_architectsRole:2016} & Cognitive distance & Full \\
\cline{1-1}
Architect-as-a-coach~\cite{Tamburri-etal_architectsRole:2016} & &\\
\cline{1-1}
Cross-functional and community-wide review of code, designs, and operations procedures~\cite{Tamburri-etal_architectsRole:2016} & &\\
\cline{1-1}
Designing an architecture on the basis of the perceived cognitive distance~\cite{Tamburri-etal_architectsRole:2016} & &\\
\cline{1-1}
Professional communication intermediaries~\cite{Tamburri-etal_architectsRole:2016} & &\\
\cline{1-1}
Stimulating engagement in newcomers, and designing architecture to accommodate buddy pairing~\cite{Tamburri-etal_architectsRole:2016} 
& &\\
\hline

Expectation management, and managers and architects using knowledge brokers to disseminate and oversee expectations~\cite{Tamburri-etal_architectsRole:2016} & Cookbook development & Full \\
\cline{1-1}
Using agile methods: in particular, employing the protecting and interface role of Scrum master~\cite{Tamburri-etal_architectsRole:2016} &  &\\
\hline

Community-wide engagement, including operations staff, clients, and user panels~\cite{Tamburri-etal_architectsRole:2016} & DevOps clash & Full \\
\cline{1-1}
Knowledge brokers and a new service or community coordinator for efficient working~\cite{Tamburri-etal_architectsRole:2016} & &\\
\cline{1-1}
Nearshoring~\cite{Tamburri-etal_architectsRole:2016} & &\\
\cline{1-1}
Standardization of software design and implementation, then separation between development and operations, and then operations offshoring~\cite{Tamburri-etal_architectsRole:2016} &
& \\
\hline

Designing a people-oriented rather than feature-oriented architecture~\cite{Tamburri-etal_architectsRole:2016} & Disengagement & Full \\
\cline{1-1}
DevOps shift-left approach to address earlier development issues and having operations staff also act as developers~\cite{Tamburri-etal_architectsRole:2016} & &\\
\hline

Architect-as-a-coach~\cite{Tamburri-etal_architectsRole:2016} & Hyper community & Full \\
\cline{1-1}
Inciting doubt through discussion and reverse logic~\cite{Tamburri-etal_architectsRole:2016} &  &\\
\hline

Daily stand-ups, and keeping track of actions, agreements, and expectations through work-division artifacts~\cite{Tamburri-etal_architectsRole:2016} & Institutional isomorphism & Full \\
\cline{1-1}
Open communication and informality, and model-based mediation of knowledge~\cite{Tamburri-etal_architectsRole:2016} & &\\
\hline

 \multirow{3}{*}{Architecture board~\cite{Tamburri_DAHLIA:2019}} &
 Invisible architecting & \multirow{3}{*}{Partial} \\
 & Lonesome architecting &  \\
 & Obfuscated architecting & \\
\hline

Full circle~\cite{Tamburri-etal_socialDebtSEinsights:2015} & Leftover techie & Partial \\
\hline

Mentoring~\cite{Catolino_etal_RefactoringCommunitySmells:2020} & Lone wolf & Full \\
\cline{1-1}
Restructure the community~\cite{Catolino_etal_RefactoringCommunitySmells:2020} & &\\
\hline

Restructure the community ~\cite{Catolino_etal_RefactoringCommunitySmells:2020} & Organizational silo & Full \\
\cline{1-1}
Create communication plan~\cite{Catolino_etal_RefactoringCommunitySmells:2020} & &\\ \cline{1-1}
Mentoring~\cite{Catolino_etal_RefactoringCommunitySmells:2020} & &\\
\hline

Architect-as-a-coach~\cite{Tamburri-etal_architectsRole:2016} & Newbie free-riding & Full \\
\cline{1-1}
Coordination management and engagement creation~\cite{Tamburri-etal_architectsRole:2016} & &\\
\cline{1-1}
Explicit empowerment of architecture decision changes~\cite{Tamburri-etal_architectsRole:2016} & &\\
\cline{1-1}
Organizational monitoring—for example, anonymous mood polling~\cite{Tamburri-etal_architectsRole:2016} & &\\
\hline
Harmonizing responsibilities, and using Scrum and agile methods~\cite{Tamburri-etal_architectsRole:2016} & Priggish members & Full \\
\hline

Culture conveyors~\cite{Tamburri-etal_socialDebtSEinsights:2015} & Prima donnas & Partial\\
& Sharing villainy & \\
\hline

Community-based contingency planning~\cite{Tamburri-etal_socialDebtSEinsights:2015} & Prima donnas & Full\\
& Solution defiance & \\
\hline

Cohesion Exercising~\cite{Catolino_etal_RefactoringCommunitySmells:2020} & Radio silence or Bottleneck & Full\\
\cline{1-1}
Mentoring~\cite{Catolino_etal_RefactoringCommunitySmells:2020} & & \\
\cline{1-1}
Create communication plan~\cite{Catolino_etal_RefactoringCommunitySmells:2020} & & \\
\cline{1-1}

Learning community~\cite{Tamburri-etal_socialDebtSEinsights:2015} & & Partial \\
\hline

Architect-as-a-coach~\cite{Tamburri-etal_architectsRole:2016} & Time warp & Full \\
\cline{1-1}
Better syncing of coordination and communication of architecture decisions and better time estimation or evidence-based scheduling~\cite{Tamburri-etal_architectsRole:2016} & &\\
\cline{1-1}
Devoting more resources to risk engineering~\cite{Tamburri-etal_architectsRole:2016} & &\\
\hline

Ad hoc individual training for the more experienced developers~\cite{Tamburri-etal_architectsRole:2016} & Unlearning & Full \\
\cline{1-1}
More active organizational approaches and resources for architecture knowledge sharing~\cite{Tamburri-etal_architectsRole:2016} & &
		\label{longtable:table_organizationalStrategiesDescriptions}
	\end{longtable}
	\footnotetext[1]{The strategy was being implemented when the study finished the data analysis.}
\end{small}

According to our findings, the team leaders who design and implement organizational strategies to mitigate community smells and pay back social debt should possess an appropriate background.
In addition to project and team management skills, leaders should have experience making socio-technical decisions and tracking their impact.
They should be familiar with social debt and community smells to identify them in practice.
They also need a deep understanding of the standards, polices, and procedures of the organization.
These skills will help leaders to build more effective organizational strategies, avoid side effects, and adjust the work environment, if needed.

\subsubsection{Frameworks}

We identified two frameworks that address social debt in software development settings.

The first framework is the Socio-technical Quality framework~\cite{Tamburri-etal_exploringCSsOS:2019}.
The framework consists of 40 socio-technical quality factors along with their metrics that researchers extracted from the literature.
These metrics help define stability thresholds where the occurrences of community smells do not increases over time.
The thresholds allow development teams to predict and mitigate the effects of the community smells.
For example, one of the 40 metrics is the number of developers or community members.
The metric indicates that community smells increase as long as the number of community members is higher than 50 and less than 200.

Further work by Catolino, et. al, defined thresholds to predict the variability of community smells~\cite{Palomba_etal_UnderstandingCSsVariability:2021}.
The socio-technical factors helped determine how much a community smell increases or decreases in software development teams over time.
Also, Palomba assessed the performance of within- and cross-project models for community smell prediction based on the 40 socio-technical quality factors~\cite{Palomba_Tamburri_PredictingEmergenceCommunitySmellSTMetrics:2021}.
According to the results, in case of a lack of historical data of projects, cross-project models represent a promising solution, yet they still need improvement.
Additional research defined statistical prediction models using metrics from this Socio-technical Quality framework~\cite{Palomba-etal_beyondTechnicalAspects:2018, Catolino-etal_genderDiversityWomen:2019}.
The next section discusses the prediction models.

Second, the DAHLIA framework quantifies social debt attributes related to the dissemination of software architecture decisions to teammates.
DAHLIA contains of four metrics that systematically measure problems with communicating architecture decisions and estimate the social debt in monetary terms.
Although DAHLIA's metrics resulted from empirical data, this theoretical framework still requires more empirical validation~\cite{Tamburri_DAHLIA:2019}.

The two frameworks described above may help reveal the causes of community smells and anticipate the effects that lead to social debt.
The preventive feature of these frameworks contrasts with the corrective nature of the organizational strategies, which appear only after a problem exists.
Although the two frameworks are potentially helpful, they need more empirical research to generalize benefits and refinement.

\subsubsection{Models}

We identified a series of models that assess and represent community smells and social debt.
In the following subsections, we define two groups of models: statistical models and social network models.

\subsubsection*{\textit{Statistical models}}

A code smells intensity prediction model used community smells to predict the future intensity of code smells with high accuracy~\cite{Palomba-etal_beyondTechnicalAspects:2018}.
This prediction model consisted of three sub-models.
The first includes only technical factors.
The second adds community smells.
The third adds socio-technical factors.
The study evaluated four community smells, \textit{Black cloud}, \textit{Lone wolf}, \textit{Organizational silo}, and \textit{Radio silence} and five code smells, long method, feature envy, blob class, spaghetti code, and misplace class.
Examples of technical and socio-technical factors used in this study are lines of code, total commits, truck-factor, and developers turnover.
The models related the factors and developers involved in community smells who modified classes to explain their relationships with code smells intensity.

The models revealed that community smells have more effect on the occurrence of code smells than socio-technical factors do.
However, each community smell was highly significant in explaining the variation of intensity of specific code smells.
For example, \textit{Black cloud} and \textit{Organizational silo} were significant predictors for long method, while \textit{Lone wolf} was significant for feature envy and spaghetti code.

To predict bug-prone code components, researchers developed seven bug prediction models~\cite{Eken_etal_EmpiricalStudyEffectCommunitySmellsBugPrediction:2021}.
This study compared community smells with process metrics as method for bug prediction.
Researchers used CodeFace4Smells to extract data and build the models based on code smells (e.g., God class and Long method) and community smells (e.g., \textit{Organizational} silo and \textit{Radio silence}).
The models also include a code smell intensity metric (i.e., the severity of code smells) and structural code metrics (e.g., lines of code and the number of methods).
Researchers built bug prediction models containing community smells, while other models excluded the community smells for comparison.
The models showed no significant differences in terms of recall and F-measure.
Nonetheless, the contribution of the community smells is statistically significant in terms of AUC.
Also, combining smell-related information to build prediction models does not significantly improve the models' performance in predicting bugs in software components.
The authors found that both code and community smells are good indicators in bug prediction models, but the performance of the models depends on project features and validation strategies.

The third model assessed the impact of gender diversity on four community smells~\cite{Catolino-etal_genderDiversityWomen:2019}.
The statistical model consisted of four sub-models, i.e., one for each community smell and the socio-technical factors related to that smell.
The target community smells were the same for the code smells intensity prediction model described above.
However, this model included more socio-technical factors, including the number of women in teams, committers or contributors, team size, and project age.

The results showed community smell instances are lower in gender-diverse teams.
Women play a significant role in reducing community smells that affect communication and coordination.
More specifically, the participation of women was significant in mitigating \textit{Black cloud} and \textit{Radio silence}.
However, gender diversity did not affect the occurrence of \textit{Lone wolf} and \textit{Organizational silo}.
The researchers performed a follow-up study ~\cite{Catolino_etal_GenderDiversityCommunitySmellsInsightsTrenches:2019} to provide more understanding of the potential gender diversity has in mitigating the community smells.
In this study, software engineering professionals perceived gender diversity as less essential than team size and professional experience in mitigating the community smells.
Despite the results, some professionals still found gender diversity important for enhancing their teams' culture, social interaction, and communication.
Although the study results contrast, more studies can be conducted to learn about gender diversity in the software industry.

Almarimi et al. implemented csDetector, a machine learning-based detection approach~\cite{Almarimi_etal_LearningDetectCommunitySmellsOSSP:2020}.
This automated approach learns and classifies seven of the nine community smells defined by Tamburri et al.~\cite{Tamburri-etal_socialDebtSEinsights:2015}, \textit{Organizational silo}, \textit{Prima donnas}, \textit{Sharing villainy}, \textit{Organizational skirmish}, \textit{Radio silence}, \textit{Black cloud}, and \textit{Solution defiance}.
The csDetector obtained an average AUC of 0.94 and achieved an average accuracy of 96\% in detecting the target community smells on 74 open-source projects.
The model also outperformed CodeFace4smell with a high detection accuracy from 94\% to 98\%.
Additionally, the model identifies influential metrics (e.g., the ratio of commits per time zone and developers per community) for improving community smell detection, such as \textit{Organizational silo} and \textit{Prima donnas}.

Almarimi et al. also conducted follow-up studies to reformulate the community smells detection.
A study proposed smell detection as a multi-label learning problem~\cite{Almarimi_etal_DetectionCommunitySmellsGPEClassifierChain:2020}.
The approach identified all target community smells with an average accuracy of 89\% and outperformed nine multi-label learning techniques that rely on different meta-algorithms and underlying learning algorithms.
Based on the performance of both models, researchers evaluated csDetector on 143 open-source projects from
GitHub~\cite{Almarimi_etal_CsDetector:2021}.
The model achieved high performance with an average accuracy of 84\% on the community smells.
Although the accuracy was lower than in the previous study~\cite{Almarimi_etal_LearningDetectCommunitySmellsOSSP:2020}, the average AUC was similar, i.e., 0.93.

Huang et al. developed and enhanced the performance of a prediction model that detects community smells based on sentiment analysis in three scenarios, i.e., cross-project, within-project, and time-wise validation~\cite{Huang_etal_PredictingCommunitySmellsOccurrenceSentiments:2021,Huang_etal_CommunitySmellOccurrencePredictionMulti-Granularity:2022}.
The study included ten process metrics to capture developers' activities.
The model predicts the impact of \textit{Organizational silo}, \textit{Radio silence}, and \textit{Lone wolf} on software developers.
It also predicts whether a developer quits their development communities due to the impact of the smells.
As a result, the model predicted the occurrence of the target community smells on developers in most cases.
It achieves mean F-measures ranging from 73\% to 92\% in the three scenarios.
Also, process metrics improve the AUC-ROC of smelly quitter prediction by 19\% to 37\%.
Moreover, community smells affect impolite developers, those with a higher workload, and those who hardly communicate.
Besides recommending the model to refactor community smells, the authors plan to integrate more process metrics, and community smells to strengthen the model.

Like frameworks, the statistical models also work as preventive management approaches.
The statistical models predict and quantitatively explain the impact of community smells and social debt on several aspects of software development settings.
Although the research contexts restrict the implementations and performance of each model, these limitations provide motivation for further empirical evaluations.
For instance, Eken et al.~\cite{Eken_etal_EmpiricalStudyEffectCommunitySmellsBugPrediction:2021} developed models that include two community smells.
The models proposed by Palomba et al.~\cite{Palomba-etal_beyondTechnicalAspects:2018}, Catolino et al.~\cite{Catolino-etal_genderDiversityWomen:2019}, and Almarimi et al.~\cite{Almarimi_etal_LearningDetectCommunitySmellsOSSP:2020, Almarimi_etal_DetectionCommunitySmellsGPEClassifierChain:2020} detect more community smells.
By including more types of community smells, the models may identify more instances of community smells depending on project features, which can influence the models' performance metrics.
Further empirical evaluations may include the rest of the community smells reported in this SLR in different research settings.
Also, the limitations are understandable since the research progress in community smells and social debt is recent.

\subsubsection*{\textit{Social-Network-based models}}\label{sec:SNBmoldels}

Social Network Analysis (SNA) can help explain the social interactions among individuals in software teams.
SNA also helps visualize scenarios where community smells affect the quality of communication, collaboration, and coordination in a software development network.

An empirical study used SNA to construct formal definitions of four community smells frequently observed in practice, \textit{Black cloud}, \textit{Lone wolf}, \textit{Organizational silo}, and \textit{Radio silence}.
The formal definitions of the four community smells helped define four developer social networks to represent the effects of the four community smells on communication and collaboration interactions~\cite{Tamburri-etal_exploringCSsOS:2019}.

In a second study, researchers used weak-ties hypotheses to represent the dependency between decision-makers and dependent teammates.
This strategy also visualizes the efficiency of the communication and explains the impact of community smells on making and sharing software architecture decisions.
For example, teammates with weak ties require longer to learn about decisions.
This work resulted in the DAHLIA framework that provides a rough estimate of social debt in monetary terms.
However, the DAHLIA framework's implementation only focuses on community smells that affect the dissemination of software architecture decisions across the software development teams~\cite{Tamburri_DAHLIA:2019}.

\subsubsection{Tools}

The literature describes two tools that provide a beginning for automated and more efficient management of community smells and social debt.
The tools have been evaluated in real settings.
The functionalities of these tools include recognizing, predicting, and monitoring the community smell effects that lead to social debt.

First, CodeFace4Smells, a deployed tool, identifies community smells in open-source projects.
Its current capability is limited to four community smells: \textit{Black cloud}, \textit{Lone wolf}, \textit{Organizational silo}, and \textit{Radio silence}~\cite{Tamburri-etal_exploringCSsOS:2019}.
As discussed in section~\ref{sec:SNBmoldels}, researchers formally defined the effects of these four community smells on communication and collaboration through developer social networks.
The tool analyzes contributors' commit history to determine the effects of community smells on collaboration.
Regarding communication, the tool mines communication from the mailing lists.
CodeFace4Smells has shown high accuracy in detecting the effects of community smells on communication and collaboration over the implemented developer social networks.
The tool also supported research on community smells and their relationships with code smells intensity~\cite{Palomba-etal_beyondTechnicalAspects:2018} and gender-diverse teams~\cite{Catolino-etal_genderDiversityWomen:2019}.

Second, YOSHI, a deployed tool, extracts information from project repositories to identify types of software engineering communities by detecting and measuring six characteristics: cohesion, formality, geodispersion, longevity, people's engagement, and structure or developers' network.
YOSHI compares the measurements to empirical thresholds from ethnographical research to determine the characteristic levels (i.e., high or low) and the types of communities~\cite{Tamburri-etal_discoveringCommunityPatterns:2019}.
Organizations can use the results produced by YOSHI to monitor the six characteristics and identify potential problems with developers' social and organizational relationships.
For example, YOSHI can use access control to repositories to establish the level of formality.
A high levels of formality can lead to community smells like \textit{Radio silence}.
YOSHI needs more empirical research to better understand the relationships between its characteristics and community smells.

The tools implement various results from social debt research.
These tools cover a limited set of community smells, so researchers need to extend them to be more comprehensive.
In addition, the tools need further development and evaluation in diverse settings.
Finally, the tools need usability evaluation.

\subsubsection{Guidelines}

The literature contained empirically-based guidelines.
These guidelines provide recommendations for managing team composition based on features of teams or communities~\cite{Tamburri-etal_architectsRole:2016}.

The guideline provides a matrix to support the management of software development teams or communities.
One axis shows 13 types of software engineering communities, while the other axis shows related features.
The matrix contains the level of relevance of each feature to every community type.
Team leaders can use this matrix to define the most suitable team composition based on projects' needs
The primary benefit is that well-defined teams increase their synergy and reduce the impact of community smells.
The matrix can also help strengthen task coordination to achieve socio-technical congruence.

\subsection{RQ4: What are the properties of the community smells?}

The software engineering literature reported 30 recurring community smells.
The literature also provides research results about the characteristics, behavior, and impact of the community smells in practice.
Based on our analysis of the literature, we identified the following 11 properties that characterize all 30 community smells:

\begin{enumerate}

\item \textbf{Occur with agile methods}: These community smells emerge as a result of the use of agile methods~\cite{Tamburri-etal_socialDebtSEinsights:2015,Tamburri_DAHLIA:2019, Palomba-etal_beyondTechnicalAspects:2018,Tamburri_splicingCSASmells:2019}.

\item \textbf{Create a debtor relationship}: The people who make poor socio-techinical decisions are the \textit{debtors}.
The wrong decisions must be corrected before their effects get worse over time (i.e. debt interest).
The danger occurs when the accumulated interest on the debt take the form of hazards, e.g., software failures~\cite{Tamburri-etal_whatSDinSE:2013,Tamburri-etal_socialDebtSEinsights:2015, Tamburri_DAHLIA:2019}.

\item \textbf{Dynamic}: Community smells emerge in organizational scenarios where people perform different roles during the software development process~\cite{Tamburri-etal_socialDebtSEinsights:2015, Tamburri-etal_architectsRole:2016,Tamburri_DAHLIA:2019,Palomba-etal_beyondTechnicalAspects:2018, Tamburri-etal_exploringCSsOS:2019,Tamburri_splicingCSASmells:2019}.
Every organizational scenario consists of particular elements (e.g., policies, procedures, and standards) that influence people's psychological state.
At the same time, behaviors can arise and change based on to the organizational climate or perceived work settings.
Thus, community smells are dynamic because of people's behavior and their perceptions of work settings.

\item \textbf{Harmful}: The effects of community smells are harmful when they are prevalent in teams and organizations.
Over time teams experience tense social interactions, lower productivity, and poor software quality~\cite{Tamburri-etal_socialDebtSEinsights:2015, Tamburri-etal_architectsRole:2016,Tamburri_DAHLIA:2019,Palomba-etal_beyondTechnicalAspects:2018, Tamburri-etal_exploringCSsOS:2019,Tamburri_splicingCSASmells:2019}.

\item \textbf{Latent}: The effects of community smells can take some time to manifest~\cite{Tamburri-etal_socialDebtSEinsights:2015,Tamburri-etal_architectsRole:2016,Tamburri_DAHLIA:2019,Palomba-etal_beyondTechnicalAspects:2018, Tamburri-etal_exploringCSsOS:2019}.
In other words, latency is a stage in which community smells incubate with no visible effects.

\item \textbf{Suggest the need for change}: Community smells warn leaders and teammates about adverse socio-technical situations.
However, the presence of community smells also suggests the need to reorganize teams, modify work setting conditions, or adjust the software development process~\cite{Palomba-etal_beyondTechnicalAspects:2018, Tamburri-etal_exploringCSsOS:2019, Tamburri_splicingCSASmells:2019}.
When leaders apply strategies to address community smells, they also address problems related to team composition, work settings, or software development processes~\cite{Catolino_etal_RefactoringCommunitySmells:2020}.
Examples of strategies are restructuring team composition, creating appropriate communication plans, and improving team cohesion.

\item \textbf{They are observable}: Software professionals can recognize the effects of community smells as they persist over time, including conflicts, professional misconduct, and software failures~\cite{Tamburri-etal_socialDebtSEinsights:2015, Tamburri-etal_architectsRole:2016, Tamburri_DAHLIA:2019,  Palomba-etal_beyondTechnicalAspects:2018}.

\item \textbf{They sabotage software quality}: Community smells undermine the software architecture process through decision unawareness~\cite{Tamburri-etal_architectsRole:2016,Tamburri_DAHLIA:2019} and intensifying architecture smells~\cite{Tamburri_DAHLIA:2019,Tamburri_splicingCSASmells:2019}.
Additionally, community smells affect the development process and implementation of software architecture decisions by intensifying code smells and making coordination more difficult~\cite{Tamburri-etal_socialDebtSEinsights:2015,Palomba-etal_beyondTechnicalAspects:2018}.
Consequently, by impacting the software development process's core phases, community smells lead to faulty software components and software failures.

\item \textbf{Innocuousness}: In this case the presence of a community smell triggers effects.
However, those effects are temporarily harmless until they become more frequent or prevalent~\cite{Tamburri-etal_socialDebtSEinsights:2015, Tamburri-etal_architectsRole:2016, Tamburri_DAHLIA:2019}.

\item \textbf{They exhibit unique effects}: Although multiple community smells may result from the same causes, their effects differentiate them~\cite{Tamburri-etal_socialDebtSEinsights:2015, Tamburri-etal_architectsRole:2016, Tamburri_DAHLIA:2019, Palomba-etal_beyondTechnicalAspects:2018, Tamburri-etal_exploringCSsOS:2019}.
Therefore, the combination of causes and effects creates a unique profile for each community smell.
For example, experience diversity is a cause for both the \textit{Time warp} and \textit{Cognitive distance} smells, however the effects of these smells are remarkably different.

\item \textbf{They are unpredictable}: Because each organization is different, the impact of the community smells differs.
The manifestation of community smell effects depends upon how people adapt to each context and changes in behavior.
Therefore, the specific effects of the community smells are unpredictable in a given context~\cite{ Tamburri-etal_whatSDinSE:2013, Tamburri-etal_socialDebtSEinsights:2015, Tamburri_DAHLIA:2019, Tamburri-etal_exploringCSsOS:2019, DiNitto_GEEZMO:2015}.

\end{enumerate}

\subsection{RQ5: How do community smells originate and evolve in software development settings?}

\begin{figure}[!ht]
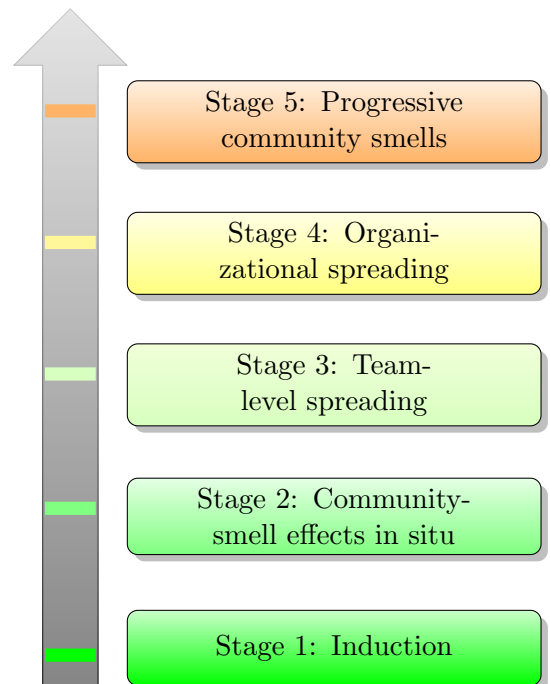

\centering
\smartdiagramset{
border color=black,
set color list={green!100, green!50, green!50!lime!25, yellow!50, orange!60}}
\smartdiagram[priority descriptive diagram]{
    Stage 1: Induction,
    Stage 2: Community-smell effects in situ,
    Stage 3: Team-level spreading,
    Stage 4: Organizational spreading,
    Stage 5: Progressive community smells}
  \caption{Community Smell Stages Framework}
  \label{fig:results_CSsStages}
\end{figure}

To answer RQ5, we developed an interpretative framework to explain the birth and evolution of community smells.
Figure~\ref{fig:results_CSsStages} shows our \textbf{Community Smell Stages Framework}
This framework depicts community smells as a chronic disease that begins in individuals then spreads across teams and organizations.
The arrow suggests that as community smells originate and evolve, the chances of prevalent effects and tangible damage increase.

\subsubsection{Stage 1: Induction}
In this stage, project managers or team leaders set up favorable conditions for the emergence of community smells.
Making and implementing socio-technical decisions in ongoing projects, e.g., changing the composition of a team, can produce a tense atmosphere.
Also, using outdated or out-of-context practices can influence software development progress.
Individuals may also begin to experience the effects of community smells in the form of irritation or stress.
Software development tasks may be at risk because teammates develop traits of professionals misbehavior, e.g., ignoring decisions.
Therefore, the causes and early effects may be undetectable at this stage.
In other words, when people begin to develop strong feelings or professional misbehavior, these effects are still invisible to managers and teammates.

\subsubsection{Stage 2: Community-smell effects in situ}

In this stage teams more frequently experience episodes of negative emotions, such as irritation, fear, or unhappiness.
This stage also includes developers overlooking procedures (e.g. neglecting to document architecture decisions) or the process of disseminating decisions~\cite{Tamburri_DAHLIA:2019}, which lowers the quality of their work.
The fact that organizations have not modified poor socio-technical decisions boosts the effects of the community smells.
However, most of these effects may still go unnoticed (i.e., team leaders are unaware of their team members' emotions or of the software development process flow) because those effects have not reached the threshold at which they become prevalent and perceptible~\cite{Tamburri-etal_exploringCSsOS:2019}.
In addition, the occurrence of these effects is unpredictable because the number of community smells fluctuates significantly compared with other periods~\cite{Tamburri-etal_exploringCSsOS:2019}.

\subsubsection{Stage 3: Team-level spreading}

In this stage, team members who experience repeated episodes of harmful emotions begin to exhibit their internal psychological states, which then affects the rest of the team.
At this point, the socio-technical decisions that triggered the community smell effects remain unchanged, resulting in the prevalence of effects throughout the team and an increase in social debt accumulation.
Consequently, these effects become negative characteristics of a team, suggesting that the team is prone to perform tasks sub-optimally and produce low-quality software artifacts.
At this stage, project managers may notice the presence of those effects.
However, developing a precise way to diagnose and repair the problems depends on the experience of the manager and the management approaches available.

At this stage, the effects of the community smells remain localized to the team.
Project managers have the opportunity to prevent potential organizational damage by regularly monitoring and mitigating the effects of community smells.
For instance, if a project manager can observe effects like conflicts and low performance, they can make changes to the situation that enables the effect thereby mitigating its potential impact.

\subsubsection{Stage 4: Organizational spreading}

In this phase, the prevalence of community smell effects leads teams to produce low-quality outcomes and to impact other teams in the organization, which may result in low-quality software.
Consequently, the quality of the processes and software products across the organization suffer.
It is more feasible for leaders to detect the effects of the community smells at this stage because their impact is evident on technical outcomes, e.g., duplicated work, faulty software components, and prototype failures.
At this point, organizations must make it a priority to allocate resources, like time and personnel, to solve the problems before the effects have external impacts, e.g. deploying a faulty software product.

\subsubsection{Stage 5: Progressive community smells}

In this stage, the effects of the community smells have an impact external to the organization~\cite{Betz-etal_sustainabilityDebt:2015}.
At this point, users experience software failures and unexpected results.
These experiences can result in lost customers and damage to the organization's reputation.
This damage can reduce the potential that new customers purchase the organization's software or services out of fear for low software quality, resulting financial losses to the organization.

\subsection{RQ6: What types of events characterize the causes and effects of community smells?}

The conceptual model for describing community smells primarily consists of  causes and effects.
We identified two types of events that can characterize the causes of community smells, i.e., \textit{independent} and \textit{dependent}.

First, \textit{independent} causes are those cause that occur naturally over time given the work setting~\cite{Tamburri-etal_socialDebtSEinsights:2015, Tamburri-etal_architectsRole:2016, Tamburri-etal_exploringCSsOS:2019}.
Based on our results, we identified 20 community smells for which there are independent events.
For example, \textit{Black cloud}, \textit{Institutional isomorphism}, and \textit{Organizational silo} were community smells whose causes occurred as independent events.

To illustrate our analysis, the causes for the \textit{Black cloud} smell are problems in the communication structure of the team.
The lack of communication protocols, skilled people, and daily information sharing meetings result from poor socio-technical decisions.
However, it is unlikely that these poor decisions led directly to the communication problems.
Rather, the poor socio-technical decisions first shaped the work environment from which the community smells emerged over time.
Therefore, we consider these causes to be independent.

Conversely, \textit{dependent} causes are those events that result from a series of events, each one making the next one more likely.
In other words, these smells result from a series of low-quality processes involving problems with sharing or communicating software architecture decisions across the teams.
We identified four community smells that fall into this category~\cite{Tamburri_DAHLIA:2019}, i.e. \textit{Architecting by osmosis}, \textit{Invisible architecting}, \textit{Lonesome architecting}, and \textit{Obfuscated architecting}.

The causes of these four community smells contain a series of sequential events associated with making and sharing architectural decisions.
In this case, when someone performs an architecture activity incorrectly, it triggers other events, which also serve as causes for community smells.
For example, \textit{Architecting by osmosis} describes a scenario where previous software architecture decisions result in software failures or inconsistencies.
Based on these failures, the customers reported problems.
This pressure from customers pushed help-desk staff and developers to make quick decisions about necessary technical changes and communicate those decisions to the architects.
Then, the architects react by modifying the architecture.
However, the architects make these decisions with incomplete information, resulting in sub-optimal decisions.

Table~\ref{table:table_CSs_TypeofEvents} summarizes the results and shows whether the causes of each smell were dependent or independent.

\begin{table}[!ht]
\caption{Characterizing the occurrence of causes based on the type of events}
\centering
\begin{threeparttable}
\begin{tabular}{ |p{5cm}|C{2.25cm}|C{2.5cm}|  }
\hline
  \multirow{3}{*}{\textbf{Community-smell name}} & \multicolumn{2}{c|}{Causes} \\
\cline{2-3}
  & \textbf{Dependent} & \textbf{Independent} \\
\hline

Architecture by osmosis & \checkmark & \\
\hline

Architecture hood & & \checkmark \\
\hline

Black cloud & & \checkmark \\
\hline

Class cognition\tnote{*} & & \\
\hline

Code red\tnote{*} & & \\
\hline

Cognitive distance & & \checkmark \\
\hline

Cookbook development & & \checkmark \\
\hline

DevOps clash & & \checkmark \\
\hline

Disengagement & & \checkmark \\
\hline

Dispersion\tnote{*} & & \\
\hline

Dissensus & & \checkmark \\
\hline

Hyper community\tnote{*} & & \\
\hline

Informality excess & & \checkmark \\
\hline

Institutional isomorphism & & \checkmark \\
\hline

Invisible architecting & \checkmark & \\
\hline

Leftover techie & & \checkmark \\
\hline

Lone wolf & & \checkmark \\
\hline

Lonesome architecting & \checkmark & \\
\hline

Newbie free-riding\tnote{*} & & \\
\hline

Obfuscated architecting & \checkmark & \\
\hline

Organizational silo & & \checkmark \\
\hline

Organizational skirmish & & \checkmark \\
\hline

Power distance & & \checkmark \\
\hline

Priggish members\tnote{*} & & \\
\hline

Prima donnas & & \checkmark \\
\hline

Radio silence or Bottleneck & & \checkmark \\
\hline

Sharing villainy & & \checkmark \\
\hline

Solution defiance & & \checkmark \\
\hline

Time warp & & \checkmark \\
\hline

Unlearning & & \checkmark \\
\hline
\end{tabular}

\begin{tablenotes}\footnotesize
\item[*] Undefined causes due to the lack of information
\end{tablenotes}
\end{threeparttable}

\label{table:table_CSs_TypeofEvents}
\end{table}

\subsection{RQ7: What are the types of causes and effects of community smells found in the literature?}

\begin{figure}[!ht]
  \centering
  \begin{subfigure}[b]{1.00\linewidth}
    \includegraphics[width=\linewidth]{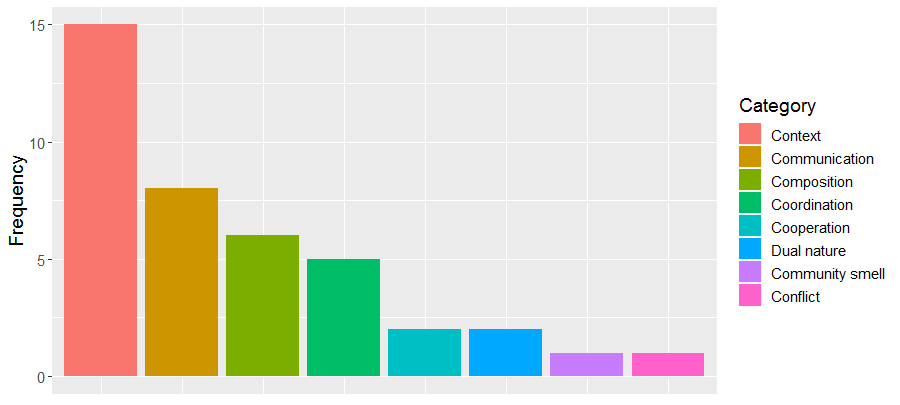}
     \caption{Types of causes}
     \label{fig:typesOfCausesCh3}
  \end{subfigure}
  \begin{subfigure}[b]{1.00\linewidth}
    \includegraphics[width=\linewidth]{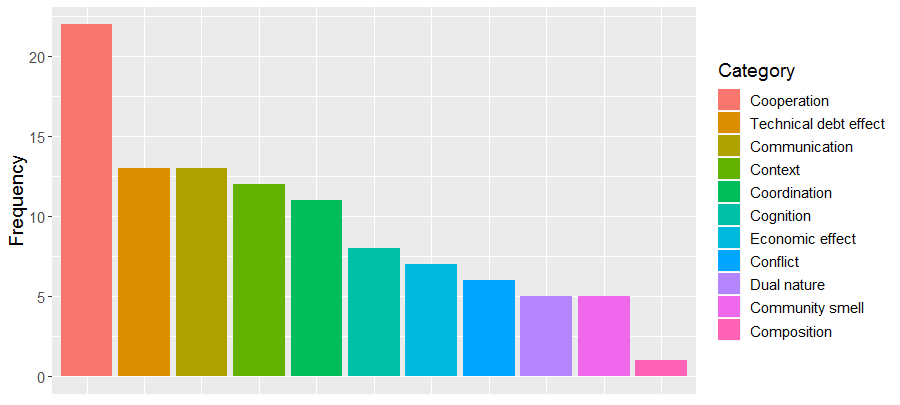}
    \caption{Types of effects}
    \label{fig:typesOfEffectsCh3}
  \end{subfigure}
  \caption{Types of causes and effects of community smells}
  \label{fig:typesOfCauses&Effects}
\end{figure}

To answer this question, we extracted a total of 44 causes and 103 effects from 30 community smells.
We conducted a coding process to analyze every cause and effect.
We used codes based on concepts from theory about the nine critical considerations for teamwork, social debt, and community smells.
To calculate code frequencies, we counted the number of times we assigned each code to each cause and effect.
The detailed results from our coding process are available online\footnote{\url{http://carver.cs.ua.edu/Data/Journals/SLR-CommunitySmells/}}.
As a result of the coding process, we identified 8 types of causes and 11 types of effects, see Figure~\ref{fig:typesOfCauses&Effects}.
We identified eight out of the nine teamwork factors, only \textit{coaching} was missing.

\subsubsection{Types of causes}

This section explains each of the types of causes.

\textbf{Context} was the most frequently occurring type of cause~\cite{Tamburri-etal_socialDebtSEinsights:2015,Tamburri-etal_architectsRole:2016}.
Some of the causes existed in organizations that used very formal processes and implemented rigid procedures or standards.
Under these conditions, teammates followed the policies to achieve team goals.
This rigid formality also shaped inflexible thinking and working methods, discouraged innovation, and increased task review time.
Other contextual causes were the geographic distribution of team members and the absence of conditions to foster knowledge sharing.
On the other hand, informal organizations exhibited different contextual features.
For example, practitioners faced a lack of protocols, misaligned organizational structures, and incompatible procedures.

\textbf{Communication} was the second most frequent type of cause~\cite{Tamburri-etal_socialDebtSEinsights:2015,Tamburri-etal_exploringCSsOS:2019,Tamburri-etal_architectsRole:2016}.
These causes related to poor communication or even the complete lack of communication.
We also found weaknesses and gaps in communication structures that affected information flow.
Some examples of these weaknesses include untraceable information sources and not sharing architecture decisions.
In terms of gaps, organizations neither provided protocols nor established space and time to share knowledge or experiences nor did they have people to distribute information to other teams.

\textbf{Composition} was the third most frequently occurring type of cause~\cite{Tamburri-etal_socialDebtSEinsights:2015,Tamburri-etal_architectsRole:2016}.
We identified causes related to team composition problems.
The causes indicated that different levels of expertise triggered conflicts in the processes for decision-making and for effectively sharing knowledge.
Also, the professional background was a problem when people who used used outdated methods could not provide innovative solutions.
In addition, changes in team composition caused older members to change their behavior and offer no help to newcomers.

\textbf{Coordination} was the next most common type of causes~\cite{Tamburri-etal_socialDebtSEinsights:2015,Tamburri-etal_architectsRole:2016}.
These causes focus on socio-technical congruence in the software development processes.
For example, software development teams struggled with high task fragmentation, unwillingness to check tasks, and wasting of time implementing modifications.
Other coordination problems include increased isolation of teammates and disengaged developers.

There were a number of less common types of causes.
\textbf{Cooperation} described situations where people avoided support or advice from their teammates~\cite{Tamburri-etal_socialDebtSEinsights:2015,Tamburri-etal_architectsRole:2016}.
It also included a lack of interest in identifying likely missing requirements that may affect the outcomes of other dependent tasks.
\textbf{Dual nature} causes included situations where the work environment or team composition included people exhibiting or practicing \textit{diverse organizational cultures}~\cite{Tamburri-etal_socialDebtSEinsights:2015,Tamburri-etal_architectsRole:2016}.
For example, the \textit{DevOps clash} smell occurs when geographically distributed teams disagree due to heterogeneity in values and standards.

\subsubsection{Types of effects}

\textbf{Cooperation} was the most frequent type of effect~\cite{Tamburri-etal_socialDebtSEinsights:2015,Tamburri-etal_architectsRole:2016, Tamburri_DAHLIA:2019,Tamburri-etal_exploringCSsOS:2019}.
Opposing teammates' attitudes led to unsuccessful cooperation across the teams.
Based on critical team-level indicators of cooperation for teamwork, this group of effects pointed out inattention to the importance of partnership, trust, and collective efficacy~\cite{Salas-etal_teamConsiderations:2015}.
Other cooperation problems like team stagnation and sharing inconsistent information across teams blurred the commitment of teams to their goals and their understanding the importance of reliable information.

\textbf{Technical debt effect} was the second most frequent type of effect~\cite{ Tamburri-etal_socialDebtSEinsights:2015, Tamburri-etal_architectsRole:2016, Tamburri_DAHLIA:2019, Palomba-etal_beyondTechnicalAspects:2018,Tamburri-etal_exploringCSsOS:2019}.
We identified 13 out of 30 community smells whose effects impacted software quality, e.g., faulty software artifacts, software failures, and duplicated code.
These results illustrate the intrinsic relationship between social debt and technical debt~\cite{Tamburri-etal_whatSDinSE:2013, Tamburri-etal_socialDebtSEinsights:2015}.

\textbf{Communication} was the third most frequent type of effect~\cite{Tamburri-etal_socialDebtSEinsights:2015,Tamburri-etal_architectsRole:2016,  Tamburri_DAHLIA:2019}.
These effects were noticeable in the structures and processes for information sharing.
For example, when the communication structures for sharing software architecture decisions excluded developers, those developers faced problems communicating with software architects.
Also, the lack of defined communication channels made sources of knowledge, like documentation, untraceable.
Regarding processes, poor decisions intensified miscommunication, delays in replying to critical requests, information overflow, and sharing confusing information.

\textbf{Context} was the the fourth most frequent type of effect~\cite{Tamburri-etal_socialDebtSEinsights:2015,Tamburri-etal_architectsRole:2016,  Tamburri_DAHLIA:2019}.
These effects included introducing changes to the task requirements, which required architects and developers to waste time performing these tasks.
Such changes also increased stress and work pressure and lowered the sense of accountability.
Developers experienced frustration due to limited communication structures.
Other effects included the lack of interest in contributing innovative solutions and people who were not in the role of architect making fast architecture decisions.
In addition, teams in reorganized software development settings made wrong assumptions about timelines to coordinate and complete activities.

\textbf{Coordination}, is the next most frequent type of effect~\cite{Tamburri-etal_socialDebtSEinsights:2015,Tamburri-etal_architectsRole:2016,  Palomba-etal_beyondTechnicalAspects:2018,Tamburri-etal_exploringCSsOS:2019}.
These problems occur when leaders faced challenges in establishing and guaranteeing socio-technical congruence in software development projects~\cite{Cataldo-etal_STC:2008}.
Overall, these effects cause development slowdowns and project delays.
For example, smells like \textit{Organizational skirmish} and \textit{DevOps clash} caused severe managerial problems and thwarted collaboration in regular operations.
These effects also included unsolved issues, wasted time, and random tasks.

\textbf{Cognition} is the next most frequent type of effects~\cite{Tamburri-etal_architectsRole:2016, Tamburri_DAHLIA:2019}.
When teams ignored the product's needs, descriptions of architectural decisions, and other software contextual information, they became lost in performing their tasks.
Also, inadequate team composition led to misinterpretation of expectations from customers and other stakeholders.
Thus, the loss of crucial knowledge across teams indicates inefficient knowledge management.

\textbf{Economic effects} is the next most frequent type of effects~\cite{Tamburri-etal_architectsRole:2016}.
As we discussed in \textit{RQ2}, software development organizations can experience economic effects in stage five.
The community smell effects are progressive and can affect customers' business operations through software failures.
Thus, the financial health of the organization is at risk due to these unsatisfied customers.

\textbf{Conflicts}, the next most frequent type of effects, occur when people with diverse professional backgrounds intensify episodes of interpersonal differences and ignoring decisions~\cite{ Tamburri-etal_socialDebtSEinsights:2015,Wurzel_etal_InterpersonalConflictsCodeReview:2022,Tamburri-etal_architectsRole:2016}.
This category of effects also includes lack of motivation and frustration because of disagreements with methods chosen for various tasks.

\textbf{Dual nature} effects are technical problems that share traits of poor task coordination and conflicts~\cite{Tamburri-etal_architectsRole:2016, Palomba-etal_beyondTechnicalAspects:2018}.
We identified situations where code smells remained unaddressed due to a lack of consensus on finding solutions and few people responsible for the maintenance of complex source code.
These effects also described a progressive loss of updated knowledge when teams lacked suitable communication structures for knowledge sharing.

\subsection{RQ8: How do the community smells affect teamwork in software development teams?}

The goal of RQ5 was to offer more insight into the poor performance of software development teams.
As teamwork is a requirement for effective team performance, we answered RQ5 in terms of the community smells that have a direct impact on teamwork factors.
Therefore, we used a portion of the results from the coding process, including the causes and effects of the 30 community smells based on eight critical factors for teamwork.

The community smells revealed leadership gaps with managing organizational resources, e.g., time and personnel.
Teammates skipped procedures or neglected generally accepted software engineering practices while conducting their tasks, which led to low-quality outcomes.
Thus, the community smell effects are characteristics of software development teams that perform poorly over time.
To contribute to the understanding of aspects determining teams' performance, we identified the need to explain more in-depth the process where community smells affect software development teams' performance.

There are nine critical factors to ensure teamwork success~\cite{Salas-etal_teamConsiderations:2015}.
By coding the causes and effects using the teamwork factors, we could determine that the causes and effects of community smells had connections to bad teamwork practices.
Based on this analysis, when community smells compromise the critical factors for effective teamwork, there is a direct impact on team performance.

\begin{figure}[!ht]
\centering
\includegraphics[width=11cm]{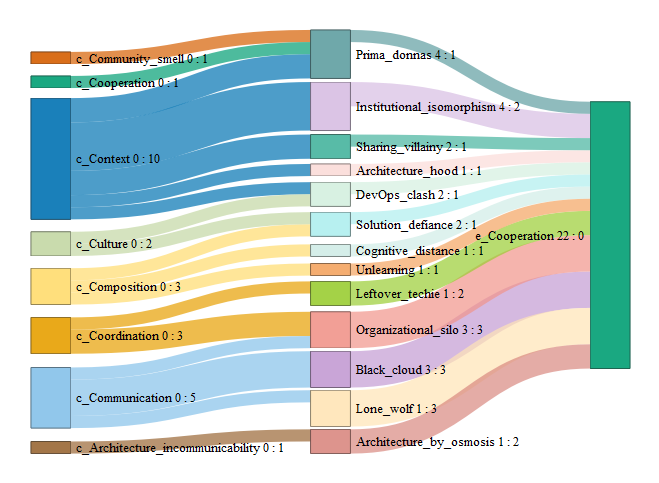}
\caption{Influence of community smells on Cooperation}
\label{fig:sankeyDiagram_CooperationCSs}
\end{figure}

\begin{figure}[!ht]
\centering
\includegraphics[width=11cm]{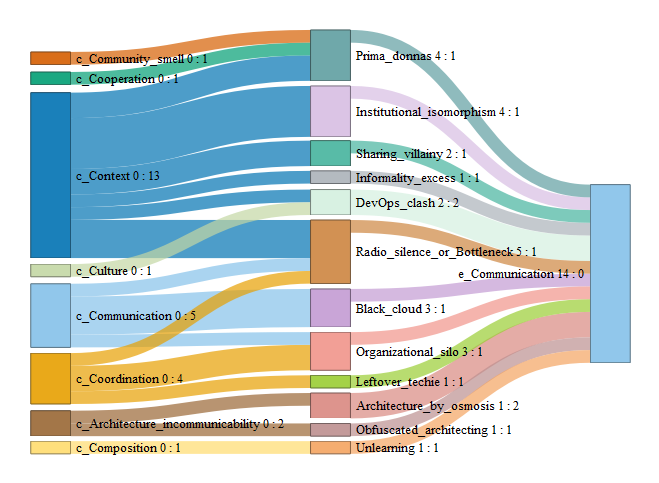}
\caption{Influence of community smells on Communication}
\label{fig:sankeyDiagram_CommunicationCSs}
\end{figure}

\begin{figure}[!ht]
\centering
\includegraphics[width=11cm]{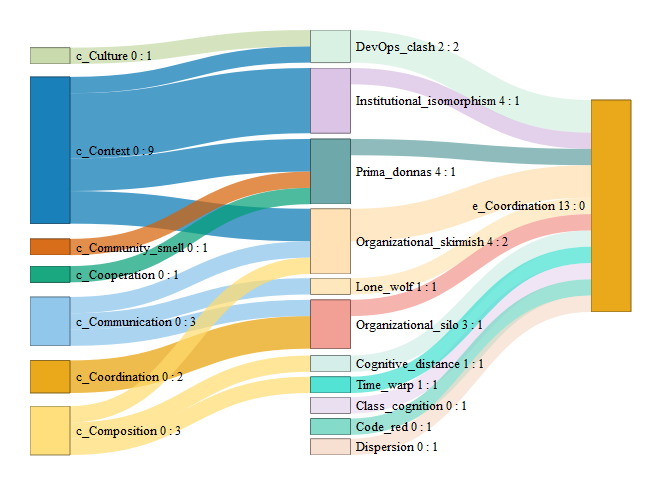}
\caption{Influence of community smells on Coordination}
\label{fig:sankeyDiagram_CoordinationCSs}
\end{figure}

Figures~\ref{fig:sankeyDiagram_CooperationCSs}, \ref{fig:sankeyDiagram_CommunicationCSs}, and \ref{fig:sankeyDiagram_CoordinationCSs} provide Sankey diagrams that show the causes and effects of the community smells on the three most common teamwork factors, \textit{cooperation}, \textit{communication}, and \textit{coordination}.
The connections at the left hand show the influence of every type of cause on the occurrence of those community smells that impact a teamwork factor.
The connections between the community smells and the teamwork factor show the effects of the community smell.
The community smell effects also lead to bad teamwork practices.
Eventually, bad teamwork practices also have an impact on team performance.
The diagrams also offer insights into the potential threat the community smells pose to each teamwork factor.
Due to space, we provide the remainder of the diagrams online\footnote{\url{http://carver.cs.ua.edu/Data/Journals/SLR-CommunitySmells/}}.

Table~\ref{longtable:table_CSsTeamworkFactors} summarizes the connection between the 30 community smells and the 8 teamwork factors.
The shaded checkmarks represent which teamwork factors are affected by which smells.
Based on this mapping, we also identified the multidimensional impact community smells have on teamwork factors
A smell can affect multiple teamwork factors.
The results indicate that 19 out of 30 community smells can affect multiple teamwork factors.
For example, \textit{Cognitive distance} and \textit{DevOps clash} can affect five of the eight teamwork factors.
Also, different community smells can affect the same factor.
\textit{Institutional isomorphism}, \textit{Organizational silo}, and \textit{Prima donnas} can affect the same teamwork factors: cooperation, communication, and coordination.

\begin{landscape}
\begin{small}
\begin{longtable}{|l|c|c|c|c|c|c|c|c|}
 \caption{Mapping overview: Influence of community smells on factors for effective teamwork\label{long_Ch3}}\\
 \hline
      \multirow{3}{*}{Community-smell name} & \multicolumn{8}{c|}{Critical Teamwork Factors}\\
\cline{2-9}
  & \multicolumn{5}{c|}{Core processes and emergent states} & \multicolumn{3}{c|}{Influencing conditions}\\
\cline{2-9}
  & Cognition & Communication & Conflict & Cooperation & Coordination & Composition & Context & Culture\\
  \hline
 \endfirsthead
\hline
 \multicolumn{9}{c}
 {{\bfseries \tablename\ \thetable{} -- continued from previous page}} \\
\hline
  \multirow{3}{*}{Community-smell name} & \multicolumn{8}{c|}{Critical Teamwork Factors}\\
\cline{2-9}
  & \multicolumn{5}{c|}{Core processes and emergent states} & \multicolumn{3}{c|}{Influencing conditions}\\
\cline{2-9}
  & Cognition & Communication & Conflict & Cooperation & Coordination & Composition & Context & Culture\\ 
 \endhead
\hline
 \multicolumn{9}{r}{{Continued on next page}} \\ \hline
\endfoot
\hline
\endlastfoot

Architecture by osmosis & & \cellcolor{black!10}\checkmark& & \cellcolor{black!10}\checkmark& & & \cellcolor{black!10}\checkmark& \\
\hline

Architecture hood & & & & \cellcolor{black!10}\checkmark& & & & \\
\hline

Black cloud & & \cellcolor{black!10}\checkmark& & \cellcolor{black!10}\checkmark& & & & \\
\hline

Class cognition & & & & & \cellcolor{black!10}\checkmark& & & \\
\hline

Code red & & & & & \cellcolor{black!10}\checkmark& & & \\
\hline

Cognitive distance & \cellcolor{black!10}\checkmark& & \cellcolor{black!10}\checkmark& \cellcolor{black!10}\checkmark& \cellcolor{black!10}\checkmark& & \cellcolor{black!10}\checkmark& \\
\hline

Cookbook development & \cellcolor{black!10}\checkmark& & & & & & & \\
\hline

DevOps clash & & \cellcolor{black!10}\checkmark& \cellcolor{black!10}\checkmark& \cellcolor{black!10}\checkmark& \cellcolor{black!10}\checkmark& & & \cellcolor{black!10}\checkmark\\
\hline

Disengagement & \cellcolor{black!10}\checkmark& & & & & & & \\
\hline

Dispersion & & & & & \cellcolor{black!10}\checkmark& \cellcolor{black!10}\checkmark& & \\
\hline

Dissensus & & & \cellcolor{black!10}\checkmark& & & & & \\
\hline

Hyper community & & & \cellcolor{black!10}\checkmark& & & & & \\
\hline

Informality excess & & \cellcolor{black!10}\checkmark& & & & & \cellcolor{black!10}\checkmark& \\
\hline

Institutional isomorphism & & \cellcolor{black!10}\checkmark& & \cellcolor{black!10}\checkmark& \cellcolor{black!10}\checkmark& & \cellcolor{black!10}\checkmark& \\
\hline

Invisible architecting & \cellcolor{black!10}\checkmark& & & & & & \cellcolor{black!10}\checkmark& \\
\hline

Leftover techie & & \cellcolor{black!10}\checkmark& & \cellcolor{black!10}\checkmark& & & & \\
\hline

Lone wolf & & & & \cellcolor{black!10}\checkmark& \cellcolor{black!10}\checkmark& & & \\
\hline

Lonesome architecting &\cellcolor{black!10}\checkmark& & & & & & \cellcolor{black!10}\checkmark& \\
\hline

Newbie free-riding & & & \cellcolor{black!10}\checkmark& & & & \cellcolor{black!10}\checkmark& \\
\hline

Obfuscated architecting & &\cellcolor{black!10}\checkmark& & & & & \cellcolor{black!10}\checkmark& \\
\hline

Organizational silo & &\cellcolor{black!10}\checkmark& & \cellcolor{black!10}\checkmark& \cellcolor{black!10}\checkmark& & \cellcolor{black!10}\checkmark& \\
\hline

Organizational skirmish & & & & & \cellcolor{black!10}\checkmark& & & \\
\hline

Power distance & & & & & & & & \\
\hline

Priggish members & & & \cellcolor{black!10}\checkmark& & & & & \\
\hline

Prima donnas & &\cellcolor{black!10}\checkmark& & \cellcolor{black!10}\checkmark& \cellcolor{black!10}\checkmark& & & \\
\hline

Radio silence or Bottleneck & &\cellcolor{black!10}\checkmark& & & & & & \\
\hline

Sharing villainy & &\cellcolor{black!10}\checkmark& & \cellcolor{black!10}\checkmark& & & & \\
\hline

Solution defiance & & & \cellcolor{black!10}\checkmark& \cellcolor{black!10}\checkmark& & & & \\
\hline

Time warp & & & & & \cellcolor{black!10}\checkmark& & \cellcolor{black!10}\checkmark& \\
\hline

Unlearning &\cellcolor{black!10}\checkmark&\cellcolor{black!10}\checkmark& & \cellcolor{black!10}\checkmark& & & & 
\label{longtable:table_CSsTeamworkFactors}
\end{longtable}
\end{small}
\end{landscape}

\section{Discussion}
    This section discusses the results of the SLR and implications for research and industry settings.
We organize this section around the key outcomes and research directions identified.

\subsection{The studies included}

We included 25 relevant studies in this SLR.
Although community smell was not the focus of some studies, researchers discussed the relationship between the main topic and community smells~\cite{Wurzel_etal_InterpersonalConflictsCodeReview:2022, Caprarelli_etal_FallaciesPitfallsDevOps:2020,Tambuerri_etal_OmniscientDevOpsAnalytics:2019, Brenner_11NontechnicalPhenomenaTD:2019, Lavallee_multiTeamsImpactSQ:2018}.
Thus, we included the studies.
We identified other studies that examined social and technical situations similar to the causes and effects of community smells~\cite{Nord-etal_agileDistress:2014,Clarke_OConnor_SituationalFacorsAffectingSD:2012, Li_BehaviorsKmeansClustering:2019, Canedo_etal_BreakingOneBarrier:2021, Etemadi_etal_TaskAssignmentTurover:2022}.
However, we excluded those studies because the authors did not establish connections between the socio-technical situations and community smells.

\subsection{Social debt, community smells, and overall research contributions}

Social debt describes dynamic and complex scenarios where daily socio-technical decisions, community smells, and software development teams play crucial roles.
The definition of social debt emphasizes the accumulation of costs generated by software development teams as a result of community smells.
However, researchers still struggle to express the effects of community smells in monetary terms.
Therefore, studies on community smells use the term \textit{effects} instead of \textit{costs} as the definition of social debt suggests.
Although the DAHLIA framework theoretically measures forms of social debt and estimates monetary costs, it is only helpful regarding architecture incommunicability~\cite{Tamburri_DAHLIA:2019}.
Thus, further research should address strategies to measure social debt items connected to the rest of the community smells.

This SLR reported five types of research contributions to manage community smells and mitigate social debt.
However, these management approaches still need more work.
For instance, the performance of prediction models and frameworks requires more empirical evaluation in different settings to generalize the results.
Additionally, researchers have built tools to examine the impact of four community smells in practice.
While these results are inspiring, there is a need to perform similar work for the remaining 26 community smells.

Regarding the tools, although CodeFace4Smells~\cite{Tamburri-etal_exploringCSsOS:2019} and YOSHI~\cite{Tamburri-etal_discoveringCommunityPatterns:2019} have performed well in empirical evaluations, they are still seen as prototypes.
Reports indicated that CodeFace4Smells and YOSHI were limited to the research context in terms of the four community smells they supported and the organizational settings.
Nevertheless, we consider these tools as advanced developments that can automatically examine repositories of software projects and identify community smells.
The authors of both tools plan to add sentiment analysis to increase performance.
This plan is consistent with the increasing emphasis on the detection of emotions connected with software artifacts~\cite{Ortu-etal_JIRARepositoryDataset:2015, Ortu-etal_emotionalSideSDJIRA:2016}.
Thus, future developments based on emotion detection can reveal community smell effects from software artifacts in the form of negative emotions.

\subsection{The 11 properties of community smells}

We identified 11 properties of community smells.
However, the distinct features of the software development settings where researchers identified the community smells can threaten the definition of these 11 properties as common denominators for all community smells.
The first feature is the \textit{organizational context}, i.e., whether the software is open-source or closed-source.
The second feature is the \textit{lifecycle phase}.
Other features that influence the occurrence and intensity of community smells are the team size~\cite{Tamburri-etal_exploringCSsOS:2019}, team structure or team composition~\cite{DeStefano_etal_SplicingCommunityPatternsSmells:2020,Tamburri-etal_discoveringCommunityPatterns:2019,Lavallee_multiTeamsImpactSQ:2018, DeStefano_etal_ImpactsSoftwareCommunityPatternsProcessProduct:2022}, and gender diversity~\cite{Catolino-etal_genderDiversityWomen:2019}.

The community smells may have several differences based on aspects of the software development settings.
Nevertheless, the smells have a similar origin, the context of action, target, and impact.
Poor socio-technical decisions are the origin of the smells.
The context of action is software development teams and organizations.
The target is people.
The impact includes low team performance, flawed processes, low software quality, and social conflicts.
Therefore, these high-level concepts build a layer where all the community smells can have the 11 properties in common.

\subsection{The \textbf{Community Smell Stages Framework}}

Our framework describes the origin and evolution of the community smells.
At each stage developers, teams, and organizations experience different manifestations of the community smells and their effects.
The framework connects the impact of the community smells with society through unsatisfied customers who are affected by software failures.
The fifth stage of the framework provides a direct link between the evolution of community smells and the transition to a social sustainability debt~\cite{Betz-etal_sustainabilityDebt:2015}.
Our framework also can serve as educational material about community smells.
For example, project managers or team leaders can use the framework as a risk-management assessment tool to help them understand the current extent of community smells in their organizations.

\subsection{The connection between the community smells and teamwork factors}

Next, the \textbf{patterns} of how the causes and effects of community smells occur and the \textbf{mapping} between the community smells and the critical factors for effective teamwork can motivate further research.
For example, researchers can apply software risk assessment~\cite{Kitchenham_EUR:1997, Dey_managinginSD:2007} to extend our work.
In this case, \textit{quantitative risk assessment} could examine the impact of such events by estimating probabilities.
This type of research would contribute to the development of more robust mitigation approaches to optimize team performance.

We identified a that the community smells have a multidimensional impact on teamwork factors.
In this scenario, when an organization attempts to mitigate one community smell, other community smells can simultaneously impact the same or different teamwork factors.
Team leaders must be aware of this multidimensional impact and address it.
Therefore, a new perspective for solutions should emphasize not only tackling specific community smells but also managing the impact of the community smells on teamwork factors and team performance.
Also, team leaders can use our mapping of community smells to teamwork factors to learn about the extent of the effects of community smells and develop organizational resources to track and enhance the performance of their software development teams.

Our mapping extended the existing interpretative framework for social debt~\cite{Tamburri-etal_socialDebtSEinsights:2015}.
We add details about where team structures become sub-optimal due to the community smells.
The mapping visualized and allowed us to explain how the community smells impact each of the critical teamwork factors.
These impacts are detrimental to team performance.
Strategies focused on managing the specific community smells that reduce the synergy between development and operation teams, e.g., \textit{Organizational skirmish} and \textit{DevOps clash}~\cite{Tambuerri_etal_OmniscientDevOpsAnalytics:2019, Caprarelli_etal_FallaciesPitfallsDevOps:2020} can make use of our results to strengthen their monitoring capabilities.

\subsection{Lessons learned}

Based on these results and our own experience with this research topic, we believe these conepts about social debt have implications for other domains beyond software engineering.
In fact, social debt can be found in any field where people work in groups or teams to develop products or services.
Thus, researchers can apply the body of knowledge on social debt and community smells from this SLR to other domains and understand how poor managerial decisions affect the welfare and performance of teams.
In that way, it is possible to explain how such decisions also impact the quality of development processes, products, services, and customers.

\section{Threats to validity}
    This section describes the threats to validity of the study and how we addressed them, where possible.

\subsection{Study selection}
To collect evidence on community smells, we searched for relevant papers in multiple digital libraries that cover different software engineering journals, conferences, and workshops.
Two out of the three authors independently applied the inclusion and exclusion criteria in two phases to reduce bias.
Given that there is not one standard venue for publishing this type of work, it is possible that we missed a study.
However, we did employ snowballing to mitigate this threat as much as possible.

\subsection{Processing collected data on community smells}
\label{sec:threats:processing}
We faced some challenges in specifying the causes and effects of community smells.
First some causes and effects seemed to combine together multiple situations.
In those cases, we split them for to ease the coding process.
Second, some community smells had only scarce information available.
The literature provided incomplete causality models for four community smells, \textit{Hyper community}, \textit{Informality excess}, \textit{Newbie free-riding}, and\textit{Priggish members}.
Four other community smells had only a brief description, \textit{Class cognition}, \textit{Code red}, and \textit{Dispersion}, and \textit{Dissensus}.
Nevertheless, the descriptions of these eight community smells had hints of their causes and effects, which allowed us to reason about possible causes and effects for some of these community smells.
Despite our effort, it we could not suggest causes for six community smells, i.e., \textit{Class cognition}, \textit{Code red}, \textit{Dispersion}, \textit{Hyper community}, \textit{Newbie free-riding}, and \textit{Priggish members}.
It is possible that, due to the limited information available, we have either incompletely or incorrectly characterized one or more of the community smells.

\subsection{Generating the codes and coding process}

Since the first author generated the initial list of codes, the second and third authors took part in the following activities to mitigate any potential bias.
As a researcher with expertise in organizational behavior and teamwork, the second author reviewed the list of codes and descriptions and reviewed the assigned codes to ensure they were appropriate.
The third author, who is an expert in software engineering and human factors, also participated in this revision.

During the coding process, we faced two challenges.
First, we had only limited information for some of the causes and effects (Section~\ref{sec:threats:processing}).
In cases where the studies did not provide either causes or effects (e.g., \textit{Newbie free-riding} and \textit{Informality excess}), we inferred the codes from studies texts.
We agreed to keep the high-level codes for those causes and effects when we lacked detailed descriptions for the fine-grained coding.
Then the three authors discussed the results and suggested final adjustments to solve disagreements and avoid bias.
Again, due to the limited information, it is possible that we inferred the causes or effects incorrectly.

In cases where we were not able to infer the causes for the community smells, we still coded any effects that were present to make the most of such a collected data.
\textit{Class cognition} and \textit{Code red} are examples of these cases.
Given the limitations in our coding process, finding other approaches to code dependent events can be a target for future work.

\section{Conclusion}
    
This paper reports on the first SLR focused on analyzing and characterizing community smells in software engineering.
We reviewed 25 studies about community smells.
We divided the goal of this SLR into eight research questions.
First, we found that social debt is a concept borrowed from social research.
The definition highlights the accumulation of socio-technical costs or recurring socio-technical effects that affect the performance of software development teams.
The potential source of social debt is community smells that affect software development teams and organizations.
The community smells can also impact customers.
We identified 30 community smells that result from 44 causes and produce 103 effects.
Different community smells show how social and technical circumstances cause TD, demonstrating the connection between social debt and TD.
Therefore, social debt can be a source for TD~\cite{YliHuumo-etal_sdTeamsManagingTD:2016}.

Among the research contributions, we identified management approaches that offer applicability for addressing the effects of community smells and repaying social debt.
We arranged the management approaches into five groups: organizational strategies, frameworks, models, tools, and guidelines.
These management approaches still need more empirical research to improve accuracy and generalize results.
However, they provide evidence of relevant progress toward addressing research challenges on social debt.

Furthermore, we built a list of 11 properties of those community smells.
The analysis of these properties is based on high-level concepts including the origin of the smell and the context in which it occurs.
We produced a novel \textit{Community Smell Stages Framework} that provides a comprehensive representation of the birth and evolution of community smells.
We also characterized the occurrence of the causes and effects of community smells based on types of events.
These results are based on information reported in the literature and still need validation through new empirical studies.

In addition, this study mapped the relationship between the community smells and critical factors for teamwork.
For the community smells, we identified 8 groups of causes and 11 groups of effects.
We also were able to generate comprehensive Sankey diagrams representing the impact of community smells on each of the teamwork factors.
These results provide insight into how the various community smells could manifest within teams.
Project leadership can use this information to help identify and remove potential socio-technical problems before they occur.

In addition, the contributions from this mapping help to extend the approaches that cluster community smells~\cite{Tamburri-etal_socialDebtSEinsights:2015}.
Our mapping results also impact the framework for social debt~\cite{Tamburri-etal_socialDebtSEinsights:2015} since
we visualized and explained how the community smells may lead to poor team performance through bad teamwork practices.
In the future, we plan to empirically validate the relationships between the community smells and teamwork reported in this study.

Because the primary goal of this SLR was to build the most comprehensive picture of the essence of community smells, our contributions have implications for software engineering researchers and professionals.
For instance, this material can provide educational material for academia or training material for industry.
Researchers can also use it as a standard reference as they conduct studies on community smells and related topics.

\bibliographystyle{model1-num-names}
\bibliography{main.bib}

\appendix
    \begin{landscape}
\section{Categories to classify the studies and research contributions}\label{appendix_A}
\begin{table}[ht!]
     \centering
\begin{tabular}{ p{2.1cm} p{4.3cm} p{11.4cm}}
 \hline
 \textbf{Aspect} & \textbf{Category} & \textbf{Description}\\
 \hline
 Paper & Empirical & When direct empirical evidence supports the research results.\\
 & Theoretical & When researchers' understanding of a topic or field supports the research results. It excludes empirical evidence.\\
 \hline
 Contribution & Advice/Implication & Recommendations driven from the authors' personal opinions\\
 & Definition & A statement on the meaning of community smells and social debt in the software engineering context\\
 & Framework & A conceptual map or method helping in analyzing and managing community smells\\
 & Guidelines & List of advises or a derived outcome from the synthesis of research results\\
 & Lessons learned & Set of outcomes directly obtained from research or industrial experience\\
 & Model & Representation of an observed reality after conceptualizing its process\\
 & Organizational strategy & A socio-technical decision focus on managing community smells in industry or open source settings\\
 & Research direction & Authors' recommendations about studying community smells in unexploited software engineering fields. It excludes implications for research or future work to extend earlier research.\\
 & Theory & The construct of cause-effect relationships between determined results, which describe a repeatable phenomenon\\
 & Tool & Technology, program, or application developed, deployed and applied to manage community smells\\
 \hline
\end{tabular}

\label{table:table_propertiesCategoriesClassificationSchemes}
\end{table}
\end{landscape}

\begin{landscape}
\section{List of codes for the coding process}\label{appendix_B}
\begin{longtable}[p]{ p{3.65cm} p{5.6cm} p{8.75cm} }
 \hline
  \textbf{Core category} & \textbf{Subcategory} & \textbf{Description/examples} \\
\hline
 \endfirsthead
\hline
 \multicolumn{3}{c}
 {{\bfseries \tablename\ \thetable{} -- continued from previous page}} \\
\hline
 \textbf{Core category} & \textbf{Subcategory} & \textbf{Description/examples} \\
\hline
 \endhead
\hline
 \multicolumn{3}{r}{{Continued on next page}} \\ \hline
\endfoot
\hline
\endlastfoot
Coaching & Coaching behavior & Coaching behaviors, like role modeling and sense-making, are positively associated with perceived team effectiveness, team productivity, and team learning.\\

Cognition & Shared understanding & The result of team member interactions, e.g., shared mental models, transactive memory systems.\\

Communication & Effective team communication & It is vital in the reduction of errors, the ability to self-adjust plans in light of teamwork breakdowns, and the acknowledgment of proper information.\\
& Team communication structure	&  It influences critical team processes since how information flows among team members can influence the team’s ability to work together and achieve goals.\\

Composition	& “Big 5” personality traits & They relate to performance in field settings i.e., extraversion, agreeableness, conscientiousness, openness to experience, emotional stability.\\
& Compositional & It is an operationalization of team composition based on the assumption of isomorphism.\\
& Compilational & It is an operationalization of team composition in which team members’ attributes interact with those from other team members to create a qualitatively different team-level.\\
& Hybrid & Combination of compositional and compilational methods.\\
& Surface-level attributes & A category of team member attributes. They are readily detectable categories (e.g., age, sex, race) and easily accessible information (e.g., reputation, role).\\
& Deep-level attributes & A category of team member attributes. They are underlying psychological characteristics (e.g., personality traits, abilities, values, attitudes).\\

Community smell & NA & A community smell identified as an effect of other community smells.\\

Conflict & Process-based conflict & Conflict regarding how to divide and delegate tasks and responsibilities among team members.\\
& Relationship-based conflict & Interpersonal differences that spark annoyance or tension among team members.\\
& Task-based conflict & Differences in viewpoints or opinions regarding how members should best execute tasks.\\

Context & External context to the team	& Influences, stimuli, or actors outside the control of the team.\\
& Internal context to the team	& Situational influences within the bounds of a team, such as the typical model of communication, nature of the team’s tasks, and the structural dependence between team members.\\
& Organizational climate & Collective agreement regarding the perception of formal and informal organizational policies, practices, and procedures.\\
& Physical context & Visible features of the working environment such as temperature, lighting, or décor.\\
& Physically distributed teams & Teams operating as virtual, distributed teams and multiteam systems, and often across national or organizational boundaries.\\
& Task context & Factors such as team or individual autonomy, uncertainty, accountability, and the resources available.\\
& Threat and stress & Teams functioning in “extreme environments,” like isolation and confinement, are susceptible to commit errors.\\

Cooperation & Collective efficacy & The collective sense of competence or perceived empowerment to control the team’s function or environment.\\
& Goal commitment & The determination to achieve team goals.\\
& Psychological safety & The shared feeling of safety within a team allowing for interpersonal risk-taking.\\
& Team/collective orientation & General preference for and belief in the importance of teamwork.\\
& Team learning orientation & Shared belief regarding the degree to which team goals are geared toward learning.\\
& Trust & The shared belief that all team members contribute as required by the role and protect the team.\\

Coordination & Explicit coordination & Team members intentionally utilize mechanisms such as planning and communication to manage interdependencies.\\
& Implicit coordination	& Team members anticipate team needs and dynamically adjust their behaviors accordingly without having to be instructed.\\

Culture & Organizational heterogeneity & It is based on cultural values and norms. This is source of conflict and process loss in terms of lacking social integration (i.e., cohesion and identity), communication, and shared meaning.\\

Economic effect & NA & Effects on employment or incomes due to a decision, event, or policy.\\

Technical debt effect & NA & Negative impact on a software artifacts, software products, and projects.

\label{longtable:table_listOfCodes}
\end{longtable}

\end{landscape}

\section{Included studies in the SLR}\label{appendix_C}
\begin{small}
\begin{longtable}[p]{ p{0.2cm} p{8.5cm} p{1cm} p{2.25cm} }
 \hline
     \textbf{\#} & \textbf{Title} & \textbf{Year} &  \textbf{Paper type} \\
\hline
 \endfirsthead
\hline
 \multicolumn{4}{c}
 {{\bfseries \tablename\ \thetable{} -- continued from previous page}} \\
\hline
  \textbf{\#} & \textbf{Title} & \textbf{Year} & \textbf{Paper type} \\ 
\hline
 \endhead
\hline
 \multicolumn{4}{r}{{Continued on next page}} \\ \hline
\endfoot
\hline
\endlastfoot

1 & Community Smell Occurrence Prediction on Multi-Granularity by Developer-Oriented Features and Process Metrics~\cite{Huang_etal_CommunitySmellOccurrencePredictionMulti-Granularity:2022} & 2022 & Empirical\\

2 & Impacts of software community patterns on process and product: An empirical study~\cite{DeStefano_etal_ImpactsSoftwareCommunityPatternsProcessProduct:2022} & 2022 & Empirical\\

3 & Interpersonal Conflicts During Code Review: Developers' Experience and Practices~\cite{Wurzel_etal_InterpersonalConflictsCodeReview:2022} & 2022 & Empirical\\

4 & Understanding community smells variability: a statistical approach~\cite{Palomba_etal_UnderstandingCSsVariability:2021} & 2021 & Empirical\\

5 & csDetector: an open source tool for community smells detection~\cite{Almarimi_etal_CsDetector:2021} & 2021 & Empirical\\

6 & Predicting Community Smells’ Occurrence on Individual Developers by Sentiments~\cite{Huang_etal_PredictingCommunitySmellsOccurrenceSentiments:2021} & 2021 & Empirical\\

7 & Predicting the emergence of community smells using socio-technical metrics: A machine-learning approach~\cite{Palomba_Tamburri_PredictingEmergenceCommunitySmellSTMetrics:2021} & 2021 & Empirical\\

8 & An empirical study on the effect of community smells on bug prediction~\cite{Eken_etal_EmpiricalStudyEffectCommunitySmellsBugPrediction:2021} & 2021 & Empirical\\

9 & On the detection of community smells using genetic programming-based ensemble classifier chain~\cite{Almarimi_etal_DetectionCommunitySmellsGPEClassifierChain:2020} & 2020 & Empirical\\

10 & Learning to detect community smells in open source software projects~\cite{Almarimi_etal_LearningDetectCommunitySmellsOSSP:2020} & 2020 & Empirical\\

11 & Refactoring Community Smells in the Wild: The Practitioner’s Field Manual~\cite{Catolino_etal_RefactoringCommunitySmells:2020} & 2020 & Empirical\\

12 & Splicing Community Patterns and Smells: A Preliminary Study~\cite{DeStefano_etal_SplicingCommunityPatternsSmells:2020} & 2020 & Empirical\\

13 & Fallacies and Pitfalls on the Road to DevOps: A Longitudinal Industrial Study~\cite{Caprarelli_etal_FallaciesPitfallsDevOps:2020} & 2020 & Empirical\\

14 & Exploring Community Smells in Open-Source: An Automated Approach~\cite{Tamburri-etal_exploringCSsOS:2019} & 2019 & Empirical\\

15 & Gender Diversity and Community Smells: Insights From the Trenches~\cite{Catolino_etal_GenderDiversityCommunitySmellsInsightsTrenches:2019} & 2019 & Empirical\\

16 & Gender Diversity and Women in Software Teams: How Do They Affect Community~\cite{Catolino-etal_genderDiversityWomen:2019} Smells? & 2019 & Empirical\\

17 & Splicing Community and Software Architecture Smells in Agile Teams: An industrial Study~\cite{Tamburri_splicingCSASmells:2019} & 2019 & Empirical\\

18 & Software Architecture Social Debt: Managing the Incommunicability Factor~\cite{Tamburri_DAHLIA:2019} & 2019 & Empirical\\

19 & Omniscient DevOps Analytics~\cite{Tambuerri_etal_OmniscientDevOpsAnalytics:2019} & 2019 & Theoretical\\

20 & Balancing Resources and Load: Eleven Nontechnical Phenomena that Contribute to Formation or Persistence of Technical Debt~\cite{Brenner_11NontechnicalPhenomenaTD:2019} & 2019 & Empirical\\

21 & Beyond Technical Aspects: How Do Community Smells Influence the Intensity of Code Smells?~\cite{Palomba-etal_beyondTechnicalAspects:2018} & 2018 & Empirical\\

22 & Discovering community patterns in open-source: a systematic approach and its evaluation~\cite{Tamburri-etal_discoveringCommunityPatterns:2019} & 2018 & Empirical\\
 
23 & Are We Working Well with Others? How the Multi Team Systems Impact Software Quality~ \cite{Lavallee_multiTeamsImpactSQ:2018} & 2018 & Empirical\\

24 & The Architect's Role in Community Shepherding~\cite{Tamburri-etal_architectsRole:2016} & 2016 & Empirical\\ 

25 & Social debt in software engineering: insights from industry~\cite{Tamburri-etal_socialDebtSEinsights:2015} & 2015 & Empirical
\label{longtable:table_includedStudiesFeatures}
\end{longtable}
\end{small}

\begin{landscape}
\section{Community smells found in the literature}\label{appendix_D}
\begin{small} 
\begin{longtable}[p]{p{5cm} p{7.5cm} p{7.5cm}}
 \hline
     \textbf{Name} & \textbf{Causes}& \textbf{Effects}\\
\hline
 \endfirsthead
\hline
 \multicolumn{3}{c}
 {{\bfseries \tablename\ \thetable{} -- continued from previous page}} \\
\hline
 \textbf{Name} & \textbf{Causes} & \textbf{Effects}\\
\hline
 \endhead
\hline
 \multicolumn{3}{r}{{Continued on next page}} \\ \hline
\endfoot
\hline
\endlastfoot

Architecture by osmosis~\cite{Tamburri_DAHLIA:2019} &
The effects of certain decisions reach clients and product operators but such decisions result in inoperable software. \newline
Product is operating and clients report many inconsistencies. \newline
Operators, pushed by clients, share malcontent with developers and suggest technical changes. \newline
Developers evaluate (and sometimes partially implement) possible technical changes and suggest change to architecture decisions. \newline
Architects make necessary changes in decisions with knowledge that was partially filtered by all communication layers in the development network.
&
Lack of vision \newline
Mistrust \newline
Decision localization \newline
Poor decision documentation \newline
Architecture erosion\\
\hline

Architecture hood~\cite{Tamburri-etal_socialDebtSEinsights:2015} &
Geographical and sociotechnical dispersion of architecture decisions &
Uncooperative behaviour across the community \newline
Solution defiance\\
\hline

Black cloud~\cite{Tamburri-etal_socialDebtSEinsights:2015, Palomba-etal_beyondTechnicalAspects:2018} &
Lack of boundary spanners \newline
Lack of sharing protocols \newline
Lack of sharing initiatives
&
Mistrust \newline
Unsanctioned initiative, e.g., people taking matters and decisions in their own hands \newline
Rise of egotistic behavior that leads to the inception of \textit{Organizational Silo} \newline
Information obfuscation\\
\hline

Class cognition~\cite{Palomba-etal_beyondTechnicalAspects:2018} &
No causes provided by or found in the literature
&
Modular structure and refactored classes are more difficult to understand and contribute for newcomers\\
\hline

Code red~\cite{Palomba-etal_beyondTechnicalAspects:2018} &
No causes provided by or found in the literature
&
Extremely complex classes that can be managed by 1-2 people at most\\
\hline

Cognitive distance~\cite{Tamburri-etal_architectsRole:2016} &
Experience diversity
&
Wasted time \newline
Wasted operations resources \newline
Lack of an optimal understanding across different operations areas \newline
Misinterpretation of expectations \newline
Pitting newbies versus experts \newline
Faulty or smelly code \newline
Additional development costs \newline
Mistrust across the development network\\
\hline

Cookbook development~\cite{Tamburri-etal_architectsRole:2016} &
Thinking in an old framework e.g, the waterfall model
&
Mismatched expectations between customers and the rest of the community\\
\hline

DevOps clash~\cite{Tamburri-etal_architectsRole:2016} &
Geographic dispersion
&
Slower development \newline
Ineffective operations \newline
The inability to bridge between different thought worlds across development and operations \newline
“Stickiness” of knowledge transfer \newline
Clashes between the development and operations cultures \newline
Increased project costs \newline
Lack of trust-building\\
\hline

Disengagement~\cite{Tamburri-etal_architectsRole:2016} &
Lack of engagement in development  \newline
Lack of curiosity
&
Missing software development contextual information  \newline
Wild assumptions\\
\hline

Dispersion~\cite{Palomba-etal_beyondTechnicalAspects:2018} &
No causes provided by or found in the literature
&
Fragmentation of a previously existing group or modularized collaboration structure in the community \newline
Haphazard work \newline
Normal maintenance activities in the community are more difficult to carry out and coordinate\\
\hline

Dissensus~\cite{Palomba-etal_beyondTechnicalAspects:2018} &
Inability to achieve consensus on how to proceed despite repeated attempts at it
&
Code smell remains as-is or teams are unable to find a common solution\\
\hline

Hyper community~\cite{Tamburri-etal_architectsRole:2016} &
No causes provided by or found in the literature
&
Increased turbulence \newline
Buggy software\\
\hline

Informality excess~\cite{Tamburri-etal_architectsRole:2016} &
Relative absence of information management and control protocols
&
Low accountability of both development and operations staff \newline
Information spillover\\
\hline

Institutional isomorphism~\cite{Tamburri-etal_architectsRole:2016} &
Excessive conformity to standards \newline
Lack of innovation \newline
Using a formal structure to achieve community goals \newline
Rigid thinking from different parts of the community
&
A negative impact on team spirit \newline
Lack of innovation \newline
Stagnation \newline
Lack of collaboration \newline
Lack of communication \newline
A less flexible or static product\\
\hline

Invisible architecting~\cite{Tamburri_DAHLIA:2019} &
Architecture decisions are made or changed rapidly \newline
Product is developed and operates as well (e.g., refactoring) \newline
Thereby, architecture documents are not used properly and/or few architects are present \newline
Also, architecture decisions are too big to implement \newline
Thus, new team is added to the development network to implement the changes
&
Decision unawareness \newline
Product version and architecture misalignment \newline
Solutions defiance \newline
Time waste\\
\hline

Leftover techie~\cite{Tamburri-etal_socialDebtSEinsights:2015} &
Increased isolation between development and operations people
&
Seemingly egotistical behaviour for knowledge and status awareness sharing \newline
General lack of trust in technicians in sharing results and current status \newline
Lack of communication or miscommunication\\
\hline

Lone wolf~\cite{Tamburri-etal_exploringCSsOS:2019, Catolino-etal_genderDiversityWomen:2019} &
Absence of communication with one of the developers who prefer working independently from the others
&
Unsanctioned architectural decisions made by contributors who carry out their work irrespective or regardless of their peers \newline
Software developers exhibiting uncooperative behaviour \newline
Software developers exhibiting mistrust \newline
Developer free-riding \newline
Side effects generated due to Organizational Silo (communication decay and negative influence on developer awareness) \newline
Delays due to Organizational Silo and Lone Wolf simultaneously \newline
Code duplication \newline
Code churn \\
\hline

Lonesome architecting~\cite{Tamburri_DAHLIA:2019} &
Architects are too few and far apart. \newline
Non architects are forced to make decisions.  \newline
Not enough time dedicated to disseminating decision and related changes
&
Decision unawareness \newline
Lack of awareness on the product’s needs \newline
Time waste \newline
Overly fast decision-making to “patch-up” \newline
Misalignment between product version and architecture \\
\hline

Newbie free-riding~\cite{Tamburri-etal_architectsRole:2016} &
No causes provided by or found in the literature
&
High work pressure \newline
Irritation \newline
Demotivation of non-senior members\\
\hline

Obfuscated architecting~\cite{Tamburri_DAHLIA:2019} &
Legacy and new product are operating together or being integrated \newline
New architecture decisions imply implementation changes that necessitate new people to be included in the development network \newline
New people do not have the needed "legacy" frame of mind
&
Single communication points for architecture decisions \newline
Sociotechnical code churn \newline
Time waste \newline
Developers frustration \\
\hline

Organizational silo~\cite{Tamburri-etal_socialDebtSEinsights:2015, Tamburri-etal_exploringCSsOS:2019} &
High decoupling between tasks \newline
Lack of communication \newline
Lack of cooperation in checking task dependencies \newline
&
Tunnel vision with a consequent lack of creativity and lack of cooperation \newline
Tunnel vision with a consequent lack of collaboration \newline
Developers make architecture decisions on their own without the necessary background and premises \newline
Developers make architecture decisions on their own using different format every time \newline
Community filled with wasted resources e.g., time \newline
Decaying communication across sub-communities and consequent negative effects on developers' situational awareness \newline
Solution defiance \newline
Duplication of code\\
\hline

Organizational skirmish~\cite{Tamburri-etal_socialDebtSEinsights:2015} &
Different communication level \newline
Different expertise level \newline
Organizational change \newline
Different business processes
&
Project delay \newline
Project failure \\
\hline

Power distance~\cite{Tamburri-etal_architectsRole:2016} &
Lack of architecture knowledge sharing
&
Additional project costs \newline
Financial loss \newline
Lost bids\\
\hline

Priggish members~\cite{Tamburri-etal_architectsRole:2016} &
No causes provided by or found in the literature
&
Additional project costs \newline
Frustrated team members\\
\hline

Prima donnas~\cite{Tamburri-etal_socialDebtSEinsights:2015} &
Innovation inertia \newline
Organizational inertia \newline
Irreceptiveness to changes/support \newline
Silo effects
&
Seemingly condescending and egotistical behaviour \newline
Lack of collaboration \newline
Lack of communication \\
\hline

Radio silence or Bottleneck~\cite{Tamburri-etal_socialDebtSEinsights:2015, Tamburri-etal_exploringCSsOS:2019} &
Highly formal and complex organizational structure \newline
Proposed changes within every software development phase require an extraordinary quantity of time to be implemented \newline
Time waste \newline
Hidden or counterintuitive information (and broker) locations \newline
Highly regularized procedures
&
Communication delays, i.e., answering critical emails or posts \\
\hline

Sharing villainy~\cite{Tamburri-etal_socialDebtSEinsights:2015} &
Lack of incentive to value knowledge sharing \newline
Lack of activities promoting useful knowledge sharing and synch \newline
&
Undefined information flow \newline
Lower engagement in the community in knowledge sharing e.g., the shared information is outdated, unconfirmed or wrong \\
\hline

Solution defiance~\cite{Tamburri-etal_socialDebtSEinsights:2015} &
Homophile groups
&
Uncooperative behavior \newline
Ignoring decisions\\
\hline

Time warp~\cite{Tamburri-etal_architectsRole:2016} &
Experience diversity
&
Low software architecture quality  \newline
Malfunctioning software or code smells \newline
Losing face in the community \newline
Unsolved operations issues \newline
Unsatisfied customers\\
\hline

Unlearning~\cite{Tamburri-etal_architectsRole:2016} &
Experience diversity
&
Lack of engagement \newline
Gradual loss of the new knowledge or best practices

 \label{longtable:table_communitySmellsCausesEffects}
\end{longtable}
\end{small}

\end{landscape}

\end{document}